\documentclass[twocolumn,english,aps,superscriptaddress,prb,floatfix,longbibliography]{revtex4-2}

\usepackage{subdepth}
\usepackage{color}
\newcommand{\cbl}{\color{blue}}

\usepackage{xcolor}
\usepackage{braket}
\usepackage{graphicx}
\usepackage{epsfig}
\usepackage{epstopdf}
\usepackage{float}
\usepackage{subfigure}
\usepackage{bbold}
\usepackage{soul}

\usepackage{dcolumn}
\usepackage{amssymb}
\usepackage{dutchcal}
\usepackage{amsmath,amsfonts,amsthm,bm}
\usepackage{wasysym}
\usepackage{bm}

\usepackage[colorlinks,bookmarks=false,citecolor=blue,linkcolor=red,urlcolor=blue]{hyperref}
\usepackage{verbatim}

\usepackage{booktabs}

\usepackage{times}

\begin{document}

\title{Signatures of fractionalization in the optical phonons of hyperhoneycomb  Kitaev  magnet $\beta$-Li$_2$IrO$_3$}
\author{Swetlana Swarup}
\affiliation{School of Physics and Astronomy, University of Minnesota, Minneapolis, MN 55455, USA}
\author{Susmita Singh}
\affiliation{School of Physics and Astronomy, University of Minnesota, Minneapolis, MN 55455, USA}
\author{Natalia B. Perkins}
\affiliation{School of Physics and Astronomy, University of Minnesota, Minneapolis, MN 55455, USA}

\date{\today}

\begin{abstract}
  In this study, we propose that the signatures of spin fractionalization in quantum magnets can be identified through a detailed analysis of the temperature dependence of the asymmetric Fano lineshape of optical phonons overlapping with a continuum of spin excitations. We focus on the hyperhoneycomb magnet $\beta$-Li$_2$IrO$_3$, a promising candidate for being in proximity to a three-dimensional Kitaev quantum spin liquid. The Raman response in $\beta$-Li$_2$IrO$_3$ notably displays a distinctive asymmetric Fano lineshape in the 24 meV Raman-active optical phonon. This asymmetry arises from the interaction between the discrete phonon mode and the spin excitation continuum, which could be fractionalized if the material is indeed near a quantum spin-liquid phase. Our theoretical model considers the coupling of this optical phonon to Majorana fermions in the Kitaev model on the hyperhoneycomb lattice. Our findings reveal that the temperature-dependent Fano lineshape is consistent with the fractionalization of spins into Majorana fermions and $\mathbb{Z}{_2}$ fluxes.
 \end{abstract}
\maketitle
\section{Introduction}

    Quantum spin liquids (QSLs) are among the most fascinating phases of matter due to their fractionalized quasiparticles and nontrivial topological properties \cite{Balents2010,savary2016,knolle2019field,Broholm2020}. 
Among the various types of QSLs, the exactly solvable Kitaev model has garnered significant attention for its potential realization in real materials 
\cite{kitaev2006,jackeli2008mott}. Numerous candidate materials have been proposed, including
 $\alpha-\text{RuCl}_3$, $\text{Na}_2\text{IrO}_3, \text{Cu}_3\text{NaIr}_2\text{O}_6$ and $\alpha$-, $\beta$-, $\gamma$-$\text{Li}_2\text{IrO}_3$, $\text{H}_3\text{LiIr}_2\text{O}_6,$ 
 to name a few \cite{trebst2022, takagi2019concept,Tsirlin2022,Rousochatzakis2024}.

However, identifying  any  QSLs, including Kitaev QSLs, in real materials poses a unique challenge because they are difficult to detect through direct experimental probes \cite{takagi2019concept, hermanns2018physics, knolle2019field}. Nevertheless, progress has been made by combining information from various probes such as specific heat measurements, inelastic neutron and Raman scattering, resonant inelastic X-ray scattering (RIXS), and thermal transport to gain insights into the different properties that characterize a QSL \cite{knolle2019field, takagi2019concept, Broholm2020}.

Recently, the study of phonon dynamics has emerged as a powerful tool to uncover the dynamics of underlying fractionalized degrees of freedom \cite{Ye2020,metavitsiadis2020phonon,feng2021temperature,Metavitsiadis2022,feng2022,feng2022sound,singh2023phonon,singh2024,vitor2024,roser2024}. In particular, it was observed that sound attenuation measured in $\alpha$-RuCl$_3$ through ultrasound experiments~\cite{hauspurg2023fractionalized} shows characteristic temperature behavior that supports the scattering of acoustic phonons from Majorana fermions in the extended Kitaev spin liquid, as predicted in earlier theoretical studies~\cite{Ye2020,metavitsiadis2020phonon,feng2021temperature,singh2023phonon}. Additionally, both theoretical~\cite{feng2022,metavitsiadis2020phonon}  and experimental studies~\cite{Sandilands2015,Glamazda2017, Li2019,Wulferding2020,Sahasrabudhe2020} have noted that optical phonons in QSL candidate materials, probed through Raman spectroscopy, exhibit a characteristic Fano lineshape. This lineshape arises from the overlap of phonons with an underlying continuum of fractionalized excitations. Its evolution with temperature and in the presence of an external magnetic field can provide valuable insights into the nature of these fractionalized excitations.

Motivated by these studies, in this paper
 we investigate the low-energy optical phonon dynamics
 in  $\beta$-Li$_2$IrO$_3$ \cite{biffin2014unconventional,Modic2014NC,takayama2015hyperhoneycomb,
 Ruiz2017,Tsirlin2022}, a promising platform for the relatively rare QSL behavior on a three-dimensional lattice with bond-anisotropic Ising-like  interactions~\cite{OBrien2016,perreault2015theory,halasz2017RIXS,eschmann2020thermodynamic}. 
  The Kitaev model on the hyperhoneycomb lattice, describing these interactions,  is exactly solvable, featuring a QSL ground state with fractionalized excitations: gapless Majorana fermions and gapped $\mathbb{Z}_2$ fluxes. The ground state is in the zero-flux sector, with a finite-temperature transition separating it from a high-temperature disordered flux state~\cite{eschmann2020thermodynamic}. Notably, the Majorana fermions exhibit a nodal line structure~\cite{OBrien2016}.
  These inherent differences  from a two-dimensional counterpart 
  lead to fundamentally different thermodynamic behavior, especially at low temperatures.

Experimentally, a Fano lineshape has indeed been  observed in one of the low-energy phonons in $\beta$-Li$_2$IrO$_3$~\cite{glamazda2016expt}.
Despite the fact that the minimal spin Hamiltonian for $\beta$-Li$_2$IrO$_3$ includes non-Kitaev interactions in addition to the Kitaev coupling ~\cite{ducatman2018magnetic,Halloran2022}, leading to its
complex incommensurate
order with counter-rotating spirals 
\cite{biffin2014unconventional,Modic2014NC,takayama2015hyperhoneycomb,Ruiz2017}, the material exhibits notable characteristics 
of  a `proximate spin-liquid' regime with signatures of long-lived fractionalized excitations, which separate the low-temperature ordered phase from the high-temperature paramagnetic regime~\cite{Ruiz2017,Ruiz2019,Ruiz2021,Halloran2022}.
For example, recent RIXS measurements~\cite{Ruiz2021} have shown that $\beta$-Li$_2$IrO$_3$ exhibits a characteristic
 continuum response extending up to at least 300 K, well above its ordering temperature of 38 K. This supports theoretical predictions~\cite{halasz2017RIXS} and allows for the formation of the Fano lineshape in low-energy phonons~\cite{glamazda2016expt}.

In this work, we theoretically investigate the temperature dependence of the parameters characterizing the Fano lineshape of a  24 meV phonon observed in Raman spectroscopy  in  $\beta$-Li$_2$IrO$_3$~\cite{glamazda2016expt}. To this end, we consider a spin-phonon Kitaev model on the hyperhoneycomb lattice and  systematically compute the Raman intensity.  Our focus is on Majorana fermionic degrees of freedom while we disregard the scattering processes from $\mathbb{Z}_2$ fluxes.  We show  that the spin-phonon coupling renormalizes phonon propagators and generates the salient Fano linshape observed in $\beta$-Li$_2$IrO$_3$.
Our numerical results for its temperature evolution give a good account of the  experimental observations ~\cite{glamazda2016expt}, suggesting the proximity of $\beta$-Li$_2$IrO$_3$ to a QSL phase.

The contents of the paper are organized as follows: We introduce the spin-lattice coupled Kitaev Model in Sec.~\ref{sec:model} with the details of the spin, phonon and spin-phonon coupled Hamiltonian presented in Sec.~\ref{sec:spin-ham}, \ref{sec:phonon-ham} and \ref{sec:sp-phCoupling} respectively.  One-loop corrections to the phonon self-energy
due to spin-phonon interaction are calculated in Sec.~\ref{sec:polbub}. 
In Sec.~\ref{sec:method}, we employ diagrammatic perturbation theory to evaluate the Raman intensity. We first obtain the pure magnetic Raman operator in Sec.~\ref{sec:spin-raman} and phonon Raman operator in Sec.~\ref{sec:phonon-raman}. Then, in  Sec.~\ref{sec:fano},  we systematically calculate the Raman intensity in the spin-phonon coupled system using the S-matrix expansion.
We discuss the results in Sec.~\ref{sec:results}, where
 we first determine the parameters of our theoretical model by fitting it to experimental data in Sec.~\ref{sec:fitparams}. Next, we compute the temperature evolution of the Fano parameters in Sec.~\ref{sec:tempevol}. Finally, we examine the dependence of these parameters on the Kitaev interaction strength in 
 Sec.~\ref{sec:jkdep}. We conclude by summarizing our findings in Sec.~\ref{sec:conclucsions}.

\section{Model}\label{sec:model}

We consider the Kitaev model on the three-dimensional hyperhoneycomb lattice, incorporating coupling to low-energy optical phonons via the magnetoelastic interaction:
\begin{equation}
\mathcal{H}=\mathcal{H}_{\text{s}}+\mathcal{H}_{\text{ph}}+\mathcal{H}_{\text{s-ph}}
\end{equation} 
Here, $\mathcal{H}_{\text{s}}$ represents the spin Hamiltonian, $\mathcal{H}_{\text{ph}}$ describes the phonon Hamiltonian, and $\mathcal{H}_{\text{s-ph}}$ captures the spin-phonon coupling.

\subsection{Spin Hamiltonian}\label{sec:spin-ham}
\begin{figure}
    \centering
    \includegraphics[width=\linewidth]{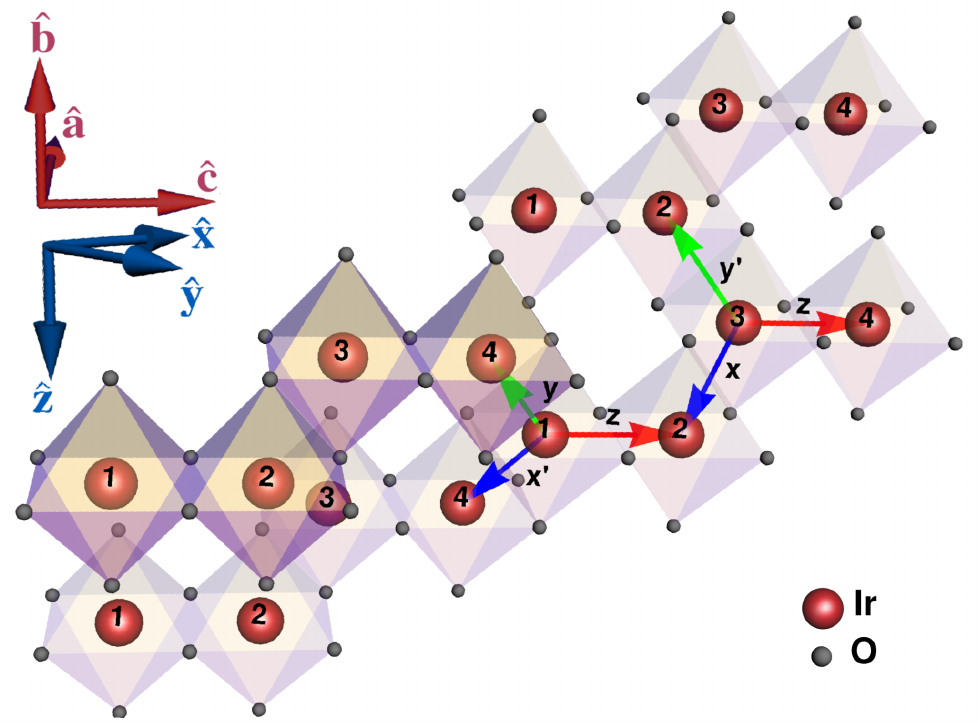}
    \caption{Nearest neighbor bonds in the hyperhoneycomb lattice. A primitive unit cell (denoted in opaque colors) consists of 4 sublattices labeled $1-4$. Four primitive unit cells make up the orthorhombic unit cell. In orthorhombic coordinates (shown in red), the bonds are $\mathbf{M_z}=(0,0,1)$, $\mathbf{M_y}=(-\frac{1}{2},\frac{1}{\sqrt{2}},-\frac{1}{2})$, $\mathbf{M_{x'}}=(\frac{1}{2},-\frac{1}{\sqrt{2}},-\frac{1}{2})$, $\mathbf{M_{y'}}=(\frac{1}{2},\frac{1}{\sqrt{2}},-\frac{1}{2})$ and $\mathbf{M_x}=(-\frac{1}{2},-\frac{1}{\sqrt{2}},-\frac{1}{2})$. The crystallographic orthorhombic coordinates are related to the Cartesian axes (shown in blue) by $\mathbf{\hat{x}}=(\mathbf{\hat{a}+\hat{c}})/\sqrt{2},\mathbf{\hat{y}}=(\mathbf{\hat{c}-\hat{a}})/\sqrt{2}$ and $\mathbf{\hat{z}=-\hat{b}}$}
    \label{fig:3dnn}
\end{figure}

The spin Hamiltonian is given by the isotropic Kitaev model on the hyperhoneycomb lattice: 
 \begin{equation}\label{Hspin}
    \mathcal{H}_{\text{s}}=-J_K \sum_t\sum_{\braket{ij}\in t} \sigma^{\alpha_t}_i\sigma^{\alpha_t}_j,\quad t \in \{x,x',y,y',z\}
\end{equation}
where $t$ represents the various nearest neighbor bonds connecting sites $i\in$ odd-sublattices and $j\in$ even-sublattices as shown in Fig. \ref{fig:3dnn} and $\alpha_t$ is the Cartesian component $\alpha_x=\alpha_{x'}=x$, \; $\alpha_y=\alpha_{y'}=y$ and $\alpha_z=z$ associated with $t$ \cite{ducatman2018magnetic}.

Using Kitaev's representation of spins  expressed by the Pauli matrices $\boldsymbol{\sigma}=(\sigma^x,\sigma^y,\sigma^z)$
in terms of Majorana fermions, $\sigma_j^{\alpha_t}=ib_j^{\alpha_t} c_j$, and defining link variables as $\hat{u}_{ij}=ib_i^{\alpha_t}b_j^{\alpha_t}$, we obtain the free Majorana fermion Hamiltonian as 
\begin{equation}\label{H-spin-free}
\mathcal{H}_{\text{s}}=i J_K\sum_t\sum_{\braket{ij}\in t} \, \hat{u}_{ij}\,c_i c_j. 
\end{equation}
We restrict to the 
zero-flux sector by setting all the link variables to +1. 
Then, the free Hamiltonian (\ref{H-spin-free}) can be easily diagonalized in Fourier space. Using the translational symmetry of the Bravais lattice and denoting the lattice point at site $i$ as ${\bf r}$, the Fourier transform is defined over the primitive unit cell as $c_{\mathbf{r},\delta}=\sqrt{\frac{2}{N}}\sum_\mathbf{k}\,e^{i\mathbf{k}\cdot\mathbf{r}}c_{\mathbf{k},\delta}$ and inversely, $c_{\mathbf{k},\delta}=\sqrt{\frac{1}{2N}}\sum_\mathbf{r}\,e^{-i\mathbf{k}\cdot\mathbf{r}}c_{\mathbf{r},\delta}$, where $\mathbf{r}=n_1\mathbf{a_1}+n_2\mathbf{a_2}+n_3\mathbf{a_3}$ and $\mathbf{k}=q_1\mathbf{k_1}+q_2\mathbf{k_2}+q_3\mathbf{k_3}$ for $n,q\in \mathbb{Z}$ and $N$ is the number of primitive unit cells. Here $\delta=1,2,3,4$ is the sublattice index in a given primitive unit cell.

The free Majorana fermion Hamiltonian in momentum-space takes the form
\begin{equation}
\mathcal{H}_{\text{s}}=\sum_\mathbf{k} C^\dagger_\mathbf{k}H_\mathbf{k}^{s}C_\mathbf{k} 
\end{equation} 
where $C_{\bf k}^\dagger=C_{\bf -k}$  with $C_{\bf k}^\dagger=(c_{\mathbf{-k},4},\, c_{\mathbf{-k},2}, \,c_{\mathbf{-k},1}, \, c_{\mathbf{-k},3})$. This arrangement is chosen such that the  matrix $H_\mathbf{k}^{s}$ takes a block diagonal form \eqref{H_matrix}.
To diagonalize this Hamiltonian, we express the Majorana fermions in terms of the complex fermions as 
\begin{equation}\label{eqn:c2a}
    C_{\bf k}=\sqrt{2} U_{\bf k} A_{\bf k}
\end{equation}
where  $U_\mathbf{k}$ is a unitary matrix that diagonalizes the matrix $H_{\bf k}^s$ and $A^\dagger_\mathbf{k}=(b_{-\mathbf{k}},\, a_{-\mathbf{k}},\, a_\mathbf{k}^\dagger ,\, b_\mathbf{k}^\dagger)$ where $a,b$ are the two flavors of the Bogoliubov complex fermions.
The spin Hamiltonian is then diagonal in the Bogoliubov basis and can be expressed as 
\begin{equation}\label{eqn:spinhambogoliu}
\mathcal{H}_{\text{s}}=\sum_\mathbf{k} A^\dagger_\mathbf{k} \Tilde{H}^{\text{s}}_\mathbf{k} A_\mathbf{k},
\end{equation}
 such that $\Tilde{H}^{\text{s}}_\mathbf{k}=2U_\mathbf{k}^\dagger H^{\text{s}}_{\mathbf{k}} U_\mathbf{k}$.  The eigenvalues are denoted by $\pm\epsilon_{\bf k}^a,\pm\epsilon_{\bf k}^b$, where $b$ represents the higher band and $a$ is the lower band as shown in Fig.~\ref{fig:ebandstruct}. Further details on the geometry of the hyperhoneycomb lattice and the explicit form of the spin Hamiltonian can be found in Appendix \ref{appendix:geometry}.

\subsection{Optical phonons}\label{sec:phonon-ham}

The free phonon Hamiltonian $\mathcal{H}_{\mathrm{ph}}\left(\mathcal{p}_i(\mathbf{r}),\mathcal{q}_i(\mathbf{r})\right)$ is a function of the canonical coordinate $\mathcal{q}_i(\mathbf{r})$,
which represents the lattice degrees of freedom in a unit cell at position ${\bf r}$,  and the corresponding canonical momentum $\mathcal{p}_i(\mathbf{r})$. In $\beta\text{-}\text{Li}_2\text{IrO}_3$, the unit cell contains
4 Iridium, 8 Lithium, and 12 Oxygen atoms.   Therefore, the canonical coordinate can be expressed as
$\mathbcal{q}(\mathbf{r})= (x_1, y_1, z_1, \ldots, x_{24}, y_{24}, z_{24})_{\bf r}$, representing the positions of the 24 atoms in the unit cell. We omit the explicit $\mathbf{r}$ dependence in the phonon displacement field because the long wavelength of the incident light leads to uniform lattice vibrations.

\begin{figure}[H]
    \centering
    \includegraphics[width=0.75\linewidth]{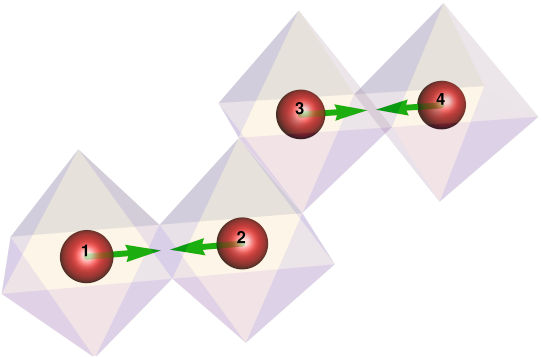}
    \caption{Visualization of the $24$ meV  $A_{g}$ optical phonon mode. We only consider the vibration of 4 Iridium atoms (shown in red) to preserve the exact solvability of the model and capture the underlying dynamics with the pure Kitaev model. The group theoretical analysis used to obtain this mode is presented in Appendix \ref{subsec:eigenmode}.}
    \label{fig:AgPhonon}
\end{figure}

We are particularly interested in the optical phonons observed in the Raman spectra at zero momentum transfer, i.e., $\mathbf{q}=0$, originating from the center of the Brillouin zone (as shown 
in Fig. \ref{fig:AgPhonon}). These modes can be described as a linear superposition of the displacement fields $u_{\Gamma}$, given by $u_{\Gamma } = \sum_{i=1}^{72} u_{\Gamma, i} \mathcal{q}_i$.  These displacement fields transform according to the irreducible representations (irreps) $\Gamma$ of the point group $D_{2h}$.
 Focusing  on the 3D vibrations of the Ir atoms only, we restrict the sum to $i=12$. The  diagonalized Hamiltonian for $\mathbf{q}=0$ phonons  with energies $\Omega_\Gamma$ can now be written as:
\begin{equation}
    \mathcal{H}_{\text{ph}}=\sum_{\Gamma}\Omega_\Gamma \bigg(\beta^\dagger_{\Gamma}\beta_{\Gamma}+\frac{1}{2}\bigg),
\end{equation} 
where $\Gamma$ labels the irreps and $\beta^\dagger_{\Gamma}(\beta_{\Gamma})$ are the phonon creation (annihilation) operators, which are related to the displacement fields as 
\begin{equation}
    u_\Gamma(t)=i\bigg(\frac{\hbar}{2\rho \delta_V \Omega_{\Gamma}}\bigg)^{\frac{1}{2}} \big( \beta_\Gamma e^{-i\Omega_\Gamma t} +\beta^\dagger_{\Gamma} e^{i\Omega_\Gamma t} \big).
\end{equation}
 Using the symmetry point group analysis, we identify the Raman-active phonon modes by considering only the vibrations of the Iridium atoms: 
\begin{equation}
    \Gamma^{\text R}_\text{Ir}=A_g\oplus B_{1g}\oplus2B_{2g}\oplus2B_{3g}.
\end{equation}
 Details  of the analysis are provided in Appendix~\ref{subsec:factorGroup}.

As mentioned in the introduction, the Fano lineshape has been observed experimentally in $\beta$-Li$_2$IrO$_3$ for the low-energy $A_g$  phonon mode centered around 24 meV~\cite{glamazda2016expt}. 
This $A_g$ phonon involves out-of-phase vibrations of Iridium atoms in the $c$-direction and stretching vibrations of the Oxygen octahedral cage~\cite{glamazda2016expt}. We visualize this mode in Fig.~\ref{fig:AgPhonon}, where we focus on the vibrations of the Iridium atoms and predominantly ignore the vibrations of the oxygen atoms, as these mainly contribute to non-Kitaev interactions. The decomposition of the phonon displacement field $u_{A_g}$ is detailed in Appendix~\ref{subsec:eigenmode}.

The bare phonon propagator in imaginary time $\tau=it$ for a given irrep $\Gamma$ is defined as $\mathcal{D}_\Gamma^{(0)}(\tau) = - \braket{T_\tau u_\Gamma(\tau)u_{\Gamma}(0)}$, where $T_\tau$ is the imaginary time ordering operator.  Performing the Fourier transform to Matsubara frequencies,
 $\mathcal{D}_\Gamma^{(0)}(i\Omega_m)=\int_0^\beta d\tau \,e^{i\Omega_m \tau}\mathcal{D}_\Gamma^{(0)}(\tau)$, we obtain
\begin{equation}
    \mathcal{D}^{(0)}_\Gamma(i\Omega_m) = -\bigg( \frac{\hbar}{\rho\delta_V} \bigg) \frac{1}{(i\Omega_m)^2-\Omega_\Gamma^2}.
\end{equation}

\subsection{Magnetoelastic coupling}\label{sec:sp-phCoupling}

 Lattice vibrations can lead to changes in bond lengths and bond angles, which in turn affect the Kitaev interaction strength $J_K$. 
Although these vibrations can potentially introduce new couplings within the system, such as Heisenberg, symmetric $\Gamma$, and antisymmetric Dzyaloshinskii-Moriya interactions, we will neglect these effects here, assuming they are small. 
Our focus will be on capturing the phonon dynamics while preserving the solvability of the Kitaev spin model.
 
 From a microscopic perspective, the coupling between the Kitaev QSL and phonon modes can be written as~\cite{Ye2020}: \begin{equation}
    \mathcal{H}_{\text{s-ph}}=-\lambda\sum_{\mathbf{r},t}\mathbf{M}_t \cdot [\mathbf{u(r)-u(r+M}_t)]\sigma_\mathbf{r}^{\alpha_t}\sigma_{\mathbf{r+M}_t}^{\alpha_t}, 
\end{equation}
where index $t$  runs over the nearest neighbor bonds  shown in Fig.~\ref{fig:3dnn} and ${\bf u(r)}$ is the lattice displacement vector at the lattice point ${\bf r}$.
By rewriting the spin bilinears in terms of Majorana fermions, we obtain a coupling between phonons and the fractionalized excitations of the Kitaev QSL.
In the Raman response, the fractionalized excitations of the QSL form a continuum, which, through the discussed magneto-elastic coupling, can result in a characteristic Fano lineshape of the optical phonons.
 As previously discussed, the Fano lineshape for the 24 meV $A_g$ phonon in $\beta$-Li$_2$IrO$_3$ has been observed, and understanding its spectral features is the focus of this study.

 We can also express
 the spin-phonon coupled Hamiltonian using the symmetry point group $D_{2h}$ in the $A_g$ irrep as 
\begin{equation}
\mathcal{H}_{\text{s-ph}}^{A_g}= u_{A_g}\big[2\lambda_{A_g}^{zz}\Sigma_{A_g}^{zz}-\lambda_{A_g}^{xx}\Sigma_{A_g}^{xx}-\lambda_{A_g}^{yy}\Sigma_{A_g}^{yy}\big],
\end{equation}
 where
\begin{equation}
    \begin{aligned}
   \Sigma_{A_g}^{xx}&=\sum_{\mathbf{r}}\bigg(\sigma_\mathbf{r}^x\sigma_\mathbf{r+M_x}^x+\sigma_\mathbf{r}^x\sigma_\mathbf{r+M_{x\prime}}^x\bigg)\\
    \Sigma_{A_g}^{yy}&=\sum_{\mathbf{r}}\bigg(\sigma_\mathbf{r}^y\sigma_\mathbf{r+M_y}^y+\sigma_\mathbf{r}^y\sigma_\mathbf{r+M_{y\prime}}^y\bigg)\\
     \Sigma_{A_g}^{zz}&=\sum_{\mathbf{r}}2\,\sigma_\mathbf{r}^z\sigma_\mathbf{r+M_z}^z.
    \end{aligned}
\end{equation}
If we assume isotropic spin-phonon coupling strengths, $\lambda_{A_g}^{xx}\sim\lambda_{A_g}^{yy}\sim\lambda_{A_g}^{zz}=\lambda_{A_g}$,  the spin-phonon coupling Hamiltonian can be   simplified as
\begin{equation}\label{HsphA1g}
    \mathcal{H}_{\text{s-ph}}^{A_g}= \lambda_{A_g}u_{A_g}\big[2\Sigma_{A_g}^{zz}-\Sigma_{A_g}^{xx}-\Sigma_{A_g}^{yy}\big].
\end{equation}
To rewrite  the spin-phonon Hamiltonian in the Bogoliubov basis, 
 we first obtain the spin-phonon coupling vertex $\Lambda_\mathbf{k}$ in the Majorana fermion representation and then perform the transformation to the Bogoliubov basis.  
  In the Majorana basis, the  spin-phonon coupling matrix in
  the momentum space reads:
\begin{equation}
    \label{eqn:lambdaK}
\Lambda_\mathbf{k}= \begin{pmatrix}
    0 & 0 & \Tilde{B}_{\mathbf{k}} & \Tilde{A}_{\mathbf{k}} \\
    0 & 0 & \Tilde{A}_{\mathbf{k}} & \Tilde{C}_{\mathbf{k}} \\
    \Tilde{B}^*_{\mathbf{k}} & \Tilde{A}^*_{\mathbf{k}} & 0 & 0 \\
    \Tilde{A}^*_{\mathbf{k}} & \Tilde{C}^*_{\mathbf{k}} & 0 & 0 
\end{pmatrix},
\end{equation}
    where 
    $\Tilde{A}_{\mathbf{k}}=-2i J_K$, $\Tilde{B}_{\mathbf{k}}=i J_K(\,e^{i\mathbf{k}\cdot\mathbf{a_2}}+\,e^{i\mathbf{k}\cdot\mathbf{a_1}})$,  and $\Tilde{C}_{\mathbf{k}}=iJ_K(1+\,e^{-i\mathbf{k}\cdot\mathbf{a_3}})$.
     Then we
use the unitary  transformation matrix  $U_{\bf k}$ from Eq.~\eqref{eqn:c2a}, and get
\begin{equation}\Tilde{\Lambda}_{\bf k}=U^\dagger_{\bf k}\Lambda_{\bf k} U_{\bf k}.
 \end{equation}
 Now, the spin-phonon Hamiltonian can be written as
\begin{equation}\label{eqn:sp-phHamAg}
\mathcal{H}_{\text{s-ph}}^{A_g}=\lambda_{A_g}\sum_\mathbf{k}A_\mathbf{k}^\dagger\Tilde{\Lambda}_\mathbf{k}A_\mathbf{k}.
\end{equation}

The spin-phonon Hamiltonian (\ref{eqn:sp-phHamAg}) does not commute with the spin Hamiltonian (\ref{eqn:spinhambogoliu}), resulting in a non-zero renormalization of the phonon propagator, as discussed in the subsequent sections.
This is a consequence of the independent basis functions  in the $A_g$ irrep of the point group $D_{2h}$ .

\subsection{Phonon Polarization Bubble}\label{sec:polbub}

\begin{figure}[t]
    \centering
    \includegraphics[width=1\linewidth]{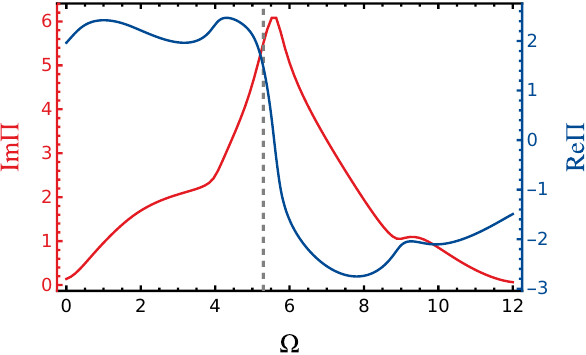}
    \caption{Plot of the real and imaginary parts of the polarization bubble $\Pi(\Omega)$, measured in units of  $\lambda_{A_g}^2$, as a function of frequency, $\Omega$ measured in units of $J_K$, computed  at  $\text{T}/J_K=0.1$. The dashed line is marking  the frequency of the optical phonon.}
    \label{fig:piplt}
\end{figure}

 To study the effects of spin-lattice coupling, we evaluate one-loop corrections to the phonon self-energy, represented by the phonon polarization bubble. 
 In the finite temperature formalism, this is determined in imaginary time as:
\begin{equation}\label{eqn:polbubexpr}
    \Pi(\tau)=\frac{\lambda_{A_g}^2}{N}\sum_\mathbf{k}\braket{(T_\tau A_\mathbf{k}^\dagger\Tilde{\Lambda}_\mathbf{k} A_\mathbf{k})(\tau)(T_\tau A_\mathbf{k}^\dagger\Tilde{\Lambda}_\mathbf{k} A_\mathbf{k})(0)},
\end{equation}
where $N$ denotes the number of units  cells. 
  Details of the derivation of phonon polarization bubble and relevant notation can be found in Appendix~\ref{Appendix:polbub}.
The final expression obtained in Matsubara space is presented below:
\begin{subequations}\label{eqn:polbubres}
\begin{align}
        \Pi^{\text{pp}}(i\Omega_m)=\frac{\lambda_{A_g}^2}{N}\sum_\mathbf{k}&\bigg(P_{aa}^{gg}\Tilde{\Lambda}_\mathbf{k}^{23}\Tilde{\Lambda}_\mathbf{k}^{32} + P_{bb}^{gg}\Tilde{\Lambda}_\mathbf{k}^{14}\Tilde{\Lambda}_\mathbf{k}^{41}
         \\&+P_{ba}^{gg}\Tilde{\Lambda}_\mathbf{k}^{13}\Tilde{\Lambda}_\mathbf{k}^{31} + P_{ab}^{gg}\Tilde{\Lambda}_\mathbf{k}^{24}\Tilde{\Lambda}_\mathbf{k}^{42}
        \bigg)\nonumber,\\
    \begin{split}\label{eqn:polbubresph}
        \Pi^{\text{ph}}(i\Omega_m)=\frac{\lambda_{A_g}^2}{N}\sum_\mathbf{k}  &P_{ba}^{g\Bar{g}}\bigg(\Tilde{\Lambda}_\mathbf{k}^{34}\Tilde{\Lambda}_\mathbf{k}^{43} +\Tilde{\Lambda}_\mathbf{k}^{12}\Tilde{\Lambda}_\mathbf{k}^{21}
        \bigg),
    \end{split}
    \end{align}
\end{subequations}
where pp- and ph- refers, correspondingly,  to 
 the particle-particle and the particle-hole  channels
  of phonon scattering from fractionalized fermionic excitations of the Kitaev spin liquid. We introduce $P_{\gamma\gamma'}^{gg}(i\Omega_m)$ and $P_{\gamma\gamma'}^{g\bar{g}}(i\Omega_m)$ as a shorthand for the convolution of Matsubara Green's functions explicitly shown in Eq.~\eqref{eqn:matsum}. For brevity, the Matsubara frequency dependence has been suppressed in the above equations. It is also worth noting that intraband processes 
   in the ph-channel only contribute to Rayleigh scattering as can be seen from the lack of $P_{aa}^{g\bar{g}}$ and $P_{bb}^{g\bar{g}}$ type terms in Eq.~\eqref{eqn:polbubresph}.
  The sum over momentum space is performed numerically, and the real and imaginary parts of the polarization bubble are plotted in 
  Fig.~\ref{fig:piplt}.

The polarization bubble renormalizes the phonon propagator via the Dyson equation as $\mathcal{D}=[(\mathcal{D}^{(0)})^{-1}-\Pi]^{-1}$, where $\Pi=\Pi'+i\, \Pi''$ is expressed in terms of its real and imaginary parts. In particular,  the phonons acquire a finite life-time given by 
the imaginary part of the renormalized phonon propagator:
\begin{equation}\label{eq:broadened_phonon}
    \mathrm{Im\, \mathcal{D}}(\Omega)\!=\!\dfrac{\lambda_{A_g}^2 \,\Pi''(\Omega_{A_g})}{\left[
    \Omega_{A_g}^2\!-\!\Omega^2\!-\!\lambda_{A_g}^2 \,\Pi'(\Omega_{A_g})\right]^2\!+\!\lambda_{A_g}^4 \,\Pi''(\Omega_{A_g})^2}.
\end{equation}
 Furthermore, as will be seen later, the phonon Raman peak parameters, such as the width, center position, and asymmetry factor, are directly related to the temperature dependence of $\Pi'$ and $\Pi''$ (Fig. \ref{fig:pitmpplt}).

\section{Methods}\label{sec:method}

In this section, we present the diagrammatic perturbation theory framework used to evaluate Raman intensity, as developed in Ref.~\cite{feng2022}.
First, we construct the vertices relevant to the scattering processes that obey the kinematic constraints. Using these vertices, we then construct the intensity diagrams and evaluate the Raman response.

\subsection{ Magnetic  Raman  operator}\label{sec:spin-raman}

The magnetic Raman  operator describes the interaction between photons and spin degrees of freedom. According to the Loudon-Fleury theory \cite{fleury1968scattering}, it can be  expressed as
\begin{equation}\label{RamanLF}
 \mathcal{R}_{\text{em-s}}=\sum_{\mu,\mu'}\varepsilon^\mu_{\text{in}}\mathcal{R}^{\mu \mu'}_{\text{em-s}}\varepsilon^{\mu'}_{\text{out}},
 \end{equation}
with
\begin{equation}
    \mathcal{R}^{\mu \mu'}_{\text{em-s}}=\nu\sum_{\mathbf{r},t}\mathbf{M}_t^\mu \mathbf{M}_t^{\mu'} \sigma_\mathbf{r}^{\alpha_t}\sigma_{\mathbf{r+M}_t}^{\alpha_t}.
\end{equation}
 Here $\boldsymbol{\varepsilon}_{\text{in}}$ and $\boldsymbol{\varepsilon}_{\text{out}}$ are the  polarization vectors of the incoming and outgoing light and $\nu$ is the photon-spin coupling constant.
$\mathcal{R}^{\mu \mu'}_{\text{em-s}}$
can also be decomposed into the irreps of the point group as follows:
\begin{equation}\mathcal{R}^{\mu \mu'}_{\text{em-s}}=\sum_{\Gamma} \alpha_\Gamma R_\Gamma^{\mu\mu'}\Sigma_{\Gamma}, 
\end{equation}
where the coefficient $\alpha_\Gamma=\frac{1}{2} \text{Tr}[R_\Gamma^T \cdot \mathcal{R}_{\text{em-s}}]$ is obtained using orthogonality of irreps, and $R_\Gamma$ is Raman tensor of irrep $\Gamma$. Here, we are interested in the magnetic Raman  operator in the  $A_g$ irrep of $D_{2h}$. Due to the independent nature of the basis functions in this irrep, the Raman operator can be written as  $\mathcal{R}^{A_g}_{\text{em-s}}=\mathcal{R}^{aa}_{\text{em-s}}+\mathcal{R}^{bb}_{\text{em-s}}+\mathcal{R}^{cc}_{\text{em-s}}$.

\subsection{ Phonon Raman operator}\label{sec:phonon-raman}
The phonon Raman  operator is determined by the interaction between the phonons and the polarization vectors of the incoming and outgoing light. 
The phonon Raman operator in the $A_g$ irrep is   given by
\begin{equation}\mathcal{R}^{A_g}_{\text{em-ph}}=\mu_{A_g} \mathcal{R}_{A_g}u_{A_g},
\end{equation}
where the Raman tensor is given by \cite{bilbaoSAM}:
\[
R_{A_g}=\begin{pmatrix}
    a & 0 & 0\\
    0 & b & 0\\
    0 & 0 & c
\end{pmatrix}
\]
For  our four-sublattice structure, the Raman tensor is simply  given by $\mathcal{R}_{A_g}=\mathcal{I}_4\otimes R_{A_g}$, where $\mathcal{I}_4$ is the $4\times4$ identity matrix.  In the following, we set  $a=b=0$, $c=1$ to consider $cc$ polarization as  experimentally reported in \cite{glamazda2016expt}.

\subsection{ Raman intensity: Fano shape of the  24 meV  $A_g$ phonon }\label{sec:fano}

Knowing the Raman operator, we can compute the Raman intensity as $I(\tau)=-
  \langle T_{\tau} \mathcal{R}({\tau})\mathcal{R}(0)\rangle,$
 where  $\langle \cdots\rangle$ is the thermal  average and $T_\tau $ is the imaginary time ordering operator. 
The effect  of  the spin-phonon coupling  modifies the Raman response, and it  can be computed systematically   
by performing the $S-$matrix expansion and expressing the coupling in the interaction picture. For finite temperatures, the intensity is given by:
\begin{equation}
 I(\tau)=-\braket{T_\tau\,\mathcal{R}(\tau)\mathcal{R}(0)e^{-\int_0^\beta d\tau'\mathcal{H}_{\text{s-ph}}(\tau')}},
 \end{equation}
 where $\mathcal{R}=\mathcal{R}_{\text{em-s}}+\mathcal{R}_{\text{em-ph}}$ and only distinct connected diagrams are summed over.  The perturbative expansion yields the following expression: 
\begin{equation}
    I(\tau)=\sum_{k=0}^\infty \, (-1)^k\,\prod_i^k\int_0^\beta d\tau_i \, \braket{T_\tau \mathcal{R}(\tau)\mathcal{R}(0)\prod_i^k\mathcal{H}^{A_g}_{\text{s-ph}}(\tau_i)}.
\end{equation}
Performing the Fourier transform to the frequency domain,
 we get:
\begin{equation}
 I(i\Omega_m) =  -\int_0^\beta  d\tau e^{i\Omega_m \tau} \,I(\tau),
 \end{equation}
where $\beta=1/T$
(with $k_B=1$ being the Boltzmann constant).

At zeroth order, $k=0$, we get the pure contribution from the magnetic and phonon Raman intensities, $I^{(0)}_{A_g}=I^{A_g}_{\text{em-ph}}+I^{A_g}_{\text{em-s}}$.
The component of the magnetic part of the intensity is given by:
 \begin{equation}
 I_{\text{em-s}}^{A_g}(i\Omega_m)=  -\int_0^\beta  d\tau e^{i\Omega_m \tau}\braket{T_{\tau}\mathcal{R}^{A_g}_{\text{em-s}}(\tau)\mathcal{R}^{A_g}_{\text{em-s}}(0)}.
 \end{equation}
 The explicit form of the Raman matrix and the intensity can be found in Appendix \ref{app:mag-raman}.
 The symmetry arguments for the  Loudon-Fleury form of the Raman operator give $I^{cc}=9I^{aa}$ and $I^{bb}=4I^{aa}$~\cite{perreault2015theory}. Thus,   we have $I_{\text{em-s}}^{A_g} \propto I^{cc}$ corresponding to the polarization geometry in the experimental setup for the Raman scattering in  $\beta$-Li$_2$IrO$_3$ \cite{glamazda2016expt}.
 The phonon contribution to the Raman intensity is calculated as:
 \begin{equation}
 I_{\text{em-ph}}^{A_g}(i\Omega_m)=  -\int_0^\beta  d\tau e^{i\Omega_m \tau}\braket{\mathcal{R}^{A_g}_{\text{em-ph}}(\tau)\mathcal{R}^{A_g}_{\text{em-ph}}(0)},
 \end{equation}
 which yields a sharply peaked delta function at the phonon frequency $\Omega_{A_g}$.

At first order, $k=1$, the  contributions to the Fano asymmetry arise. The intensity can be written as:
\begin{equation}
I_{A_g}^{(1)}(\tau) = \int_0^\beta d\tau_1 \,\braket{T_\tau\mathcal{R}(\tau)\mathcal{R}(0)\mathcal{H}^{A_g}_{\text{s-ph}}(\tau_1)}.
\end{equation}
It can be further decomposed as
\begin{equation}\label{eqn:fanoInt}
    I_{A_g}^{(1)}(\tau) =I_L^{(1)}(\tau)+I_R^{(1)}(\tau),
\end{equation}
where $I_L^{(1)}(\tau)=\int_0^\beta\, d\tau_1 \,\braket{T_\tau\mathcal{R}^{A_g}_{\text{em-ph}}(\tau)\mathcal{R}^{A_g}_{\text{em-s}}(0)\mathcal{H}^{A_g}_{\text{s-ph}}(\tau_1)}$, and $I_R^{(1)}(\tau)=\int_0^\beta\, d\tau_1 \,\braket{T_\tau\mathcal{R}^{A_g}_{\text{em-s}}(\tau)\mathcal{R}^{A_g}_{\text{em-ph}}(0)\mathcal{H}^{A_g}_{\text{s-ph}}(\tau_1)}$. These contributions are diagrammatically represented in  Eq.~\eqref{eqn:FanoDiags} where panel (a) is $I_L^{(1)}(\tau)$ and panel (b) is $I_R^{(1)}(\tau)$:

\begin{equation}\label{eqn:FanoDiags}
\begin{gathered}
    \includegraphics[width=0.65\linewidth]{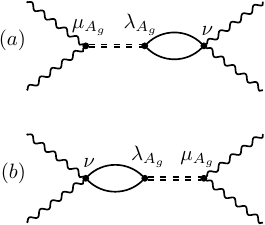}
\end{gathered}
\end{equation}
 Here, the wavy lines represent photon propagators, the dashed lines denote phonon propagators renormalized by the spin-phonon interaction, and the solid lines indicate Majorana fermion propagators. The details for the evaluation of both the left and the right diagrams are outlined in Appendix \ref{App:Fano}.
  It is also important to note that all fermionic lines in the intensity diagrams originate from Majorana fermions, as we do not account for the $\mathbb{Z}_2$ fluxes in our consideration of various scattering processes.  This is because $\mathbb{Z}_2$ fluxes are gapped, and there is a finite-temperature phase transition separating the zero-flux phase from the finite-temperature phase with flux excitations \cite{eschmann2020thermodynamic}.

Finally, we perform a Fourier transform to the Matsubara frequency, followed by an analytical continuation to real frequency, where $i \Omega_m \rightarrow \Omega + i\delta$. Here, $\Omega = \Omega_{\text{in}} - \Omega_{\text{out}}$ represents the frequency shift, defined as the difference between the incoming and outgoing photon frequencies. The total intensity is then given by:
\begin{equation}
I_{A_g} (\Omega)=I_{A_g}^{(0)} (\Omega)+I_{A_g}^{(1)}(\Omega).
\end{equation}

\section{Results}\label{sec:results}
To compare our theoretical results with experimental data, we need to fix the model parameters that appear in the total intensity. We detail this procedure in the next subsection, followed by a presentation of our findings.

\subsection{Fitting the model parameters}\label{sec:fitparams}

There are  free  coupling parameters $(\mu_{A_g},\nu,\lambda_{A_g})$ that need to be fixed before performing further quantitative computations.
First, we note that $\mu_{A_g}$ and $\nu$ control the overall intensity of the phonon Raman peak and the magnetic continuum, respectively. Since the overall magnitude of the Raman spectrum is free to rescale, increasing or reducing $\mu_{A_g}$ and $\nu$ uniformly by the same factor will retain the shape of the calculated spectrum, differing only in overall scale.
In contrast, $\lambda_{A_g}$, the spin-phonon coupling strength,   cannot be freely rescaled as it controls the phonon peak widths as well as its Fano asymmetry. Thus, fixing $\lambda_{A_g}$ with respect to $\mu_{A_g}$ and $\nu$   is crucial for accurately determining the detailed features of the overall Raman spectrum.

\begin{figure}[t]
    \centering
    \includegraphics[width=0.95\linewidth]{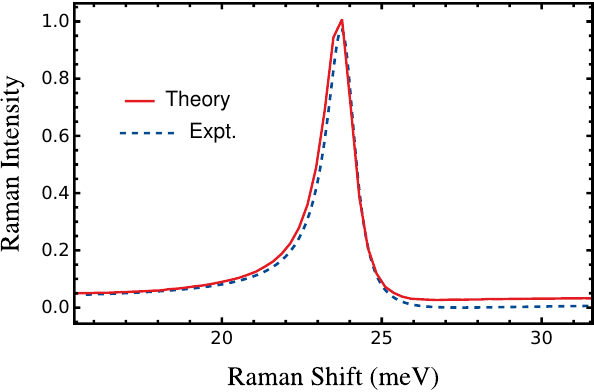}
    \caption{Plot of the experimental phonon Raman peak (dashed blue) compared with the theoretically evaluated Raman intensity (solid red) for the model parameters $\mu_{A_g}=1.0, \nu=-0.37$ and $\lambda_{A_g}=0.29$. All coupling  are in the units of $J_K$.} 
    \label{fig:modelparam}
\end{figure}

We fit these model parameters at a low temperature using the experimental data in \cite{glamazda2016expt}. We estimate $\nu$ from the ratio of the peak to continuum reported in this study, giving
 $\nu^2/\mu^2\sim 1/6$, which suggests $\nu \sim 0.41 J_K$. To set $\lambda_{A_g}$, we consider the imaginary part of the renormalized phonon propagator, expected to be a symmetric Lorentzian depending only on $\lambda_{A_g}$. By fitting it to the experimental peak with the asymmetry manually removed, we estimate $\lambda_{A_g}=0.49 J_K$.
Then, we fix the overall scale by setting
 $\mu_{A_g}=1,$ and further fine-tune these model parameters by comparing the sum of all relevant diagrams to the experimental phonon Raman peak. A comparison between the experimental data and our calculation is shown in Fig.~\ref{fig:modelparam} with the fitted parameter values $(\mu_{A_g},\nu,\lambda_{A_g})=(1.0,-0.37,0.29)$ in units of the Kitaev strength $J_K$.

The Fano lineshape of the Raman intensity at any temperature can be phenomenologically captured by the asymmetric Lorentzian function:
\begin{equation}\label{eqn:asymLor}
    I(\Omega)=\dfrac{I_0}{q^2+1}\dfrac{\left(q+\dfrac{\Omega-\omega_{0}}{\gamma}\right)^2}{1+\left(\dfrac{\Omega-\omega_{0}}{\gamma}\right)^2}.
\end{equation}
Here, the parameters $I_0,q,\omega_{0}$, and $\gamma$ control the overall intensity, asymmetry, peak position, and peak width respectively. We present the temperature evolution of the total Raman intensity in terms of the asymmetry parameter, $1/|q|$, the peak position, $\omega_{0}$, and the peak width, $\gamma$, in the following section.

\subsection{Temperature Evolution of the Fano Peak Parameters}\label{sec:tempevol}
\begin{figure}[t]
  \centering
    \includegraphics[width=0.95\linewidth]{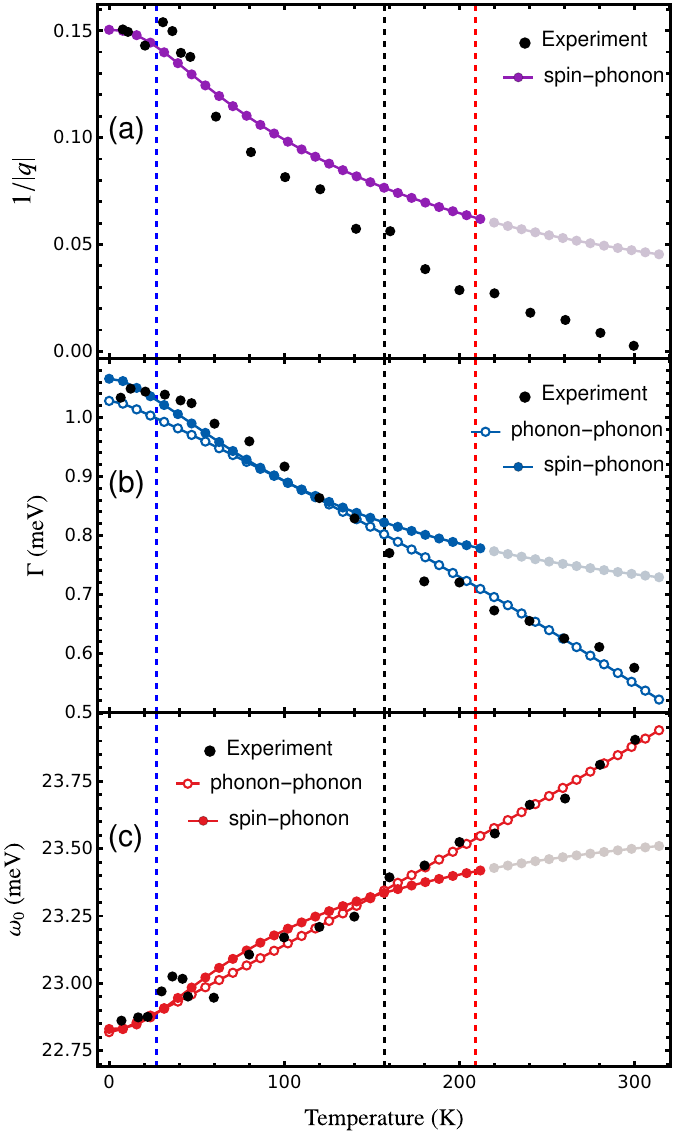}
    \caption{Theoretically computed temperature evolution of the Fano peak parameters plotted with the experimental data (shown by black circles) \cite{glamazda2016expt}. Blue ($T_L=27\,K$) and red ($T_H=209\,K$)  dashed lines mark temperature scales characteristic of the Kitaev model \cite{do2017majorana}. The spin-phonon contribution is shown in solid colors, with the grayed-out part past $T_H=209\,K$ indicating the loss of validity of the spin-phonon model as spins are no longer fractionalized.  The phonon-phonon contribution, shown in open circles, accounts for the anharmonic decay of optical phonons as described by 
    Eq.~\eqref{eqn:phononAnhar}. Phenomenological anharmonic model constants are $(A,B,C,D) = (-7.99,\,-0.0198,\,19.7,\,0.0218) \mu \mathrm{eV}$. We use Kitaev strength of $J_K=18$ meV for unit conversions.}
    \label{fig:peakparamtemp}
\end{figure}

Once the model parameters $\mu_{A_g}$, $\nu$, $\lambda_{A_g}$ are fixed we calculate the Raman intensity,  $I(\Omega)$, as a function of temperature. We then fit our calculated intensity to the asymmetric Lorentzian Eq.~\eqref{eqn:asymLor} to extract the peak parameters at each temperature. The temperature evolution of the calculated Fano peak parameters along with the experimental data \footnote{We divide by 2 the linewidth shown in temperature evolution data presented in Ref. \cite{glamazda2016expt} to obtain the correct FWHM for comparison. $\Gamma$ represents the FWHM where $\gamma$ is the HWHM such that $\Gamma=2\gamma$.} is shown in Fig.~\ref{fig:peakparamtemp}. All momentum-space integrals are performed numerically. The momentum-space is discretized into $100\times100\times100$ points. The frequency-space is discretized into $201$ points.

A note is in order here. Phonon scattering from fractionalized excitations occurs at relatively low temperatures, with the range determined by the magnitude of the Kitaev interaction. At higher temperatures, phonon-phonon scattering processes become dominant. Although we do not consider these explicitly, we  account for phonon-phonon scattering processes as described by the phenomenological anhamonicity model \cite{balkanski1983anharmonic}:
\begin{equation}\label{eqn:phononAnhar}
    \begin{aligned}
        \omega(T) = \omega_0 + C\left[1 + \frac{2}{e^{x} - 1}\right] + D\left[1 + \frac{3}{e^{y} - 1} + \frac{3}{(e^{y} - 1)^2}\right]\\
        \Gamma(T) = \Gamma_0 + A\left[1 + \frac{2}{e^{x} - 1}\right] + B\left[1 + \frac{3}{e^{y} - 1} + \frac{3}{(e^{y} - 1)^2}\right]
    \end{aligned}
\end{equation}
where $x=\hbar\omega_0/2k_B T$,  $y=\hbar\omega_0/3k_B T$ and $\omega_0$ and $\Gamma_0$ are the frequency and the linewidth of an optical mode at very low temperature, respectively. Supplementing our spin-phonon coupling model with the phonon-phonon decay at higher temperatures provides a good explanation for the experimental data.

Fig.~\ref{fig:peakparamtemp} (a) shows the temperature evolution of the Fano asymmetry, with purple circles for calculated values and black circles for experimental values. The Fano asymmetry decreases as temperature rises due to reduced pp-scattering from occupied low-energy fermionic states.

Fig.~\ref{fig:peakparamtemp} (b) shows the temperature evolution of $\Gamma$ (FWHM) of the total Raman intensity. Experimental values are shown as filled black circles, spin-phonon model calculations as filled blue circles, and anharmonic phonon-phonon model calculations as open blue circles. The decrease in $\Gamma$ with increasing temperature is due to reduced pp-scattering, leading to less broadening.

Fig.~\ref{fig:peakparamtemp} (c) shows the temperature evolution of the peak position $\omega_0$ of the total Raman intensity. Experimental values are shown as filled black circles, spin-phonon model calculations as filled red circles and anharmonic phonon-phonon model calculations as open red circles. The increase in $\omega_0$ with increasing temperature can be understood from the expression for broadened phonon peak given in Eq.~\eqref{eq:broadened_phonon}. Since the real part of the phonon polarization bubble $\Pi'$ decreases with temperature the peak of the phonon Raman intensity shifts to higher values as can be seen from the denominator of Eq.~\eqref{eq:broadened_phonon} and Fig. \ref{fig:pitmpplt}. 

As expected, the agreement between experimental and calculated values from the spin-phonon coupled model in  Fig.~\ref{fig:peakparamtemp} (a)-(c)  is better for temperatures below $T \sim 150$ K, marked by the dashed black grid-line.  
Below this temperature, the Majorana excitations of the Kitaev QSL  almost solely determine the shape of the 24 meV phonon mode through magnetoelastic coupling, resulting in a better fit with the experimental data.  However, above $T \sim 150$ K, anharmonic effects become dominant and show better agreement with the experimental data, as seen in Fig.~\ref{fig:peakparamtemp} (b), (c). Additionally, flux proliferation with increasing temperature would contribute to a further decrease in the asymmetry characteristic of the Fano lineshape.

\subsection{$J_K$ dependence of Fano asymmetry}\label{sec:jkdep}

\begin{figure}[t]
    \centering
    \includegraphics[width=0.95\linewidth]{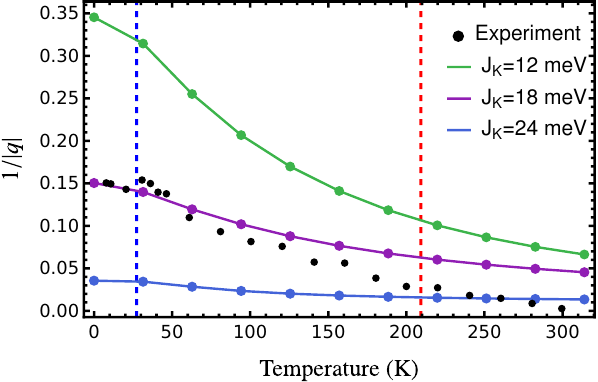}
    \caption{ The Fano parameter $1/|q|$ as a function of temperature, shown for three Kitaev couplings $J_K$ within the range of proposed Kitaev strengths for $\beta$-Li$_2$IrO$_3$. The Fano asymmetry is significantly suppressed at higher Kitaev strengths, setting a limit for the possible values of $J_K$.
    }
    \label{fig:jkvar}
\end{figure}

We now discuss the dependence of the Fano asymmetry, characterized by $1/|q|$, on the Kitaev coupling strength $J_K$, as shown in Fig.~\ref{fig:jkvar}. 
The asymmetry of the Fano lineshape is determined  by the relative position of the phonon peak and the Majorana fermion continuum. 
While the spread of the continuum scales with $J_K$, the position of phonon Raman peak depends only on the 
elastic constants. Therefore, a change in $J_K$ modifies the relative position of the phonon peak to the Majorana continuum, leading to a change in $1/|q|$. 

In Fig.~\ref{fig:jkvar}, we observe this  $J_K$ dependence of the Fano asymmetry through the temperature evolution of $1/|q|$ for three values of
$J_K=12, 18,$ and $24$ meV, shown in green, purple, and blue, respectively,with experimental data represented by filled black circles. Note that while all three curves follow the same trend of monotonically decreasing with temperature, the Fano asymmetry is suppressed more for larger values of 
 $J_K$. This trend can be understood from the aforementioned points. As $J_K$ increases the weight of the Majorana continuum shifts to the right of the phonon peak leading to smaller asymmetry. 

This dependence of the Fano asymmetry on $J_K$  can help in estimating the strength of the Kitaev coupling in 
  $\beta$-$\text{Li}_2\text{IrO}_3$. A simple comparison of the obtained curves with the experimental findings in Fig.~\ref{fig:jkvar}  suggests that 18 meV is a very plausible Kitaev coupling strength, further confirming our previous estimates from other dynamical responses \cite{ducatman2018magnetic,Rousochatzakis2018,Li2020,Li2020b,Yang2022,Halloran2022,Ruiz2021}. 

\section{Conclusion}\label{sec:conclucsions}

 In this work, we presented a theoretical investigation aimed at explaining the temperature dependence of the Fano lineshape of the 24 meV phonon peak observed by Raman spectroscopy in the three-dimensional Kitaev candidate material  $\beta$-$\text{Li}_2\text{IrO}_3$ \cite{glamazda2016expt}. We achieved this by considering a pure Kitaev model on a 3D hyperhoneycomb lattice, coupled with low-energy optical phonons.  In this model, spins fractionalize into Majorana fermions and $Z_2$ fluxes,
 However, at sufficiently low temperatures, below the flux phase transition \cite{eschmann2020thermodynamic}, the spin-phonon  interaction is between the phonons and the Majorana fermions. Therefore, we assumed that the 24 meV phonon is predominantly coupled to Majorana fermions at low temperatures. As the temperature increases and fluxes become thermally excited, the shape of the phonon peak is already mainly determined by anharmonicity effects. We expect fluxes to only add small corrections to this behavior, which we choose to disregard in the current work due to the complexity they introduce in the numerics.

We find that our spin-phonon coupling model closely captures the temperature evolution of the Fano asymmetry and other peak parameters up to the major fermionic excitation temperature of approximately 150 K. Beyond this temperature, Majorana fermions are no longer good degrees of freedom, and the scattering processes become dominated by phonon-phonon anharmonicities.  By combining these two mechanisms, we obtained  very good agreement with the experimental results \cite{glamazda2016expt}.   We also used our formalism to refine the range of the Kitaev coupling strength in $\beta$-Li$_2$IrO$_3$. By analyzing the Fano asymmetry parameter $1/|q|$, we estimated that the Kitaev coupling strength is bounded between 18 meV and 24 meV.

To conclude, our study of the Fano lineshape in Raman spectroscopy provides valuable insights into the temperature dependence of the 24 meV phonon peak in $\beta$-Li$_2$IrO$_3$ and demonstrates the potential of phonon Raman spectroscopy as a probe for other quantum spin liquids. Our theoretical approach to analyzing the Fano asymmetry and its dependence on coupling strengths, temperature, and other external parameters can be easily  extended to various spin liquid candidates. This makes it a powerful tool for extracting information about quantum spin liquid phases from Raman studies of the phonon Fano lineshape.

\section*{Acknowledgements}
The authors thank  P. Peter Stavropoulos, Vitor Dantas, Kexin Feng, Yang Yang, David Mayrhofer and Andrey Chubukov for valuable discussions. The work is supported by the U.S. Department of Energy, Office of Science, Basic Energy Sciences under Award No. DE-SC0018056.   We acknowledge the support from NSF DMR-2310318 and  the support of the Minnesota Supercomputing Institute (MSI) at the University of Minnesota. N.B.P. also acknowledges the hospitality and partial support  of the Technical University of Munich – Institute for Advanced Study and  the Alexander von Humboldt Foundation.\\

\appendix

\section{Fractionalization  in the Kitaev model on the hyperhoneycomb lattice }\label{appendix:geometry}

 In this Appendix, we review the solution of the Kitaev model on the hyperhoneycomb lattice, earlier presented in Refs.~\cite{OBrien2016,perreault2015theory,halasz2017RIXS,feng2022sound}.

\begin{figure}[h]
    \centering
    \includegraphics[width=0.95\linewidth]{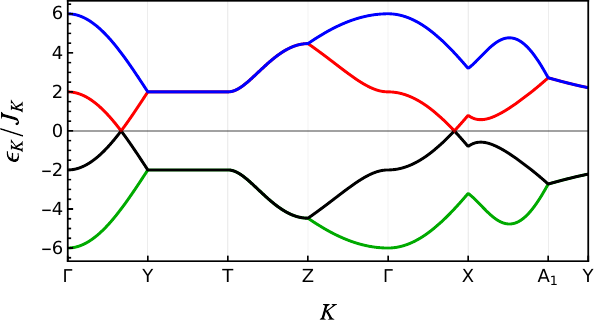}
    \caption{Band structure of the Majorana fermions on a hyperhoneycomb lattice in the zero-flux sector along the high symmetry points in the 3D Brillouin Zone \cite{SETYAWAN2010299}. The special points in the BZ are given by $\Gamma=(0,0,0)$, $Y=\left(0,\frac{2\pi}{b},0\right)$, $T=\left(0,\frac{2\pi}{b},\frac{2\pi}{c}\right)$, $Z=\left(0,0,\frac{2\pi}{c}\right)$, $X=\left(\frac{29\pi}{18a},0,0\right)$ and $A_1=\left(\frac{11\pi}{18a},\frac{2\pi}{b},0\right)$. The fermions with negative energy (bands in black and green) at momentum $\mathbf{k}$ are related to the physical fermions with positive energy (bands in blue and red) at momentum $-\mathbf{k}$ due to particle-hole symmetry.}
\label{fig:ebandstruct}
\end{figure}

The primitive unit cell of the hyperhoneycomb lattice (see Fig.~\ref{fig:3dnn}) comprises of four sublattices indexed by $\delta=1,2,3,4$. Their coordinates in the orthorhombic basis are:
\begin{eqnarray}
&&\delta_1=(0,0,0),\,
\delta_2=(0,0,\frac{c}{6}),\\
&&\delta_3=(\frac{a}{4},\frac{b}{4},\frac{c}{4}),\,
\delta_4=(\frac{a}{4},\frac{b}{4},\frac{5c}{12}).\nonumber
\end{eqnarray}
 The orthorhombic unit cell contains 
 four  primitive unit cells. The lattice vectors of the orthorhombic unit cell are 
 \begin{eqnarray}
\mathbf{a}=(a,0,0),\,
\mathbf{b}=(0,b,0),\,\mathbf{c}=(0,0,c),
\end{eqnarray}
  such that $a:b:c=1:\sqrt{2}:3$. The Cartesian axes are related to the orthorhombic axes as 
 \begin{eqnarray}
\mathbf{x}=(\mathbf{a}+\mathbf{c})/\sqrt{2},\,\mathbf{y}=(\mathbf{c}-\mathbf{a})/\sqrt{2},\, \mathbf{z}=-\mathbf{b}.
\end{eqnarray} 
 The primitive unit cell lattice vectors are then given  by
 \begin{eqnarray}
\mathbf{a_1}=(\frac{a}{2},0,\frac{c}{2}),\mathbf{a_2}=(0,\frac{b}{2},\frac{c}{2}),\mathbf{a_3}=(\frac{a}{2},\frac{b}{2},0),
\end{eqnarray} 
  so that
the reciprocal lattice vectors are 
 \begin{eqnarray}
&&\mathbf{k_1}=(\frac{2\pi}{a},-\frac{2\pi}{b},\frac{2\pi}{c}),\mathbf{k_2}=(-\frac{2\pi}{a},\frac{2\pi}{b},\frac{2\pi}{c}),\nonumber\\
&&\mathbf{k_3}=(\frac{2\pi}{a},\frac{2\pi}{b},-\frac{2\pi}{c}).
\end{eqnarray} 
In this convention, the spin-Hamiltonian can be represented as a matrix:
\begin{widetext}
        \begin{eqnarray}
\mathcal{H}_{\text{s}}=\sum_\mathbf{k} C^\dagger_\mathbf{k}H_\mathbf{k}^{s}C_\mathbf{k}=\sum_{\mathbf{k}}
\begin{pmatrix}
    c_{\mathbf{-k},4} & c_{\mathbf{-k},2} & c_{\mathbf{-k},1} & c_{\mathbf{-k},3}
\end{pmatrix} \begin{pmatrix}
    0 & 0 & B_{\mathbf{k}} & A_{\mathbf{k}} \\
    0 & 0 & A_{\mathbf{k}} & C_{\mathbf{k}} \\
    B^*_{\mathbf{k}} & A^*_{\mathbf{k}} & 0 & 0 \\
    A^*_{\mathbf{k}} & C^*_{\mathbf{k}} & 0 & 0 
\end{pmatrix} \begin{pmatrix}
    c_{\mathbf{k},4}\\c_{\mathbf{k},2}\\c_{\mathbf{k},1}\\c_{\mathbf{k},3}
\end{pmatrix},\label{H_matrix}
\end{eqnarray}
where $A_{\mathbf{k}}=-iJ_K$, $B_{\mathbf{k}}=-i J_K(\,e^{i\mathbf{k}\cdot\mathbf{a_2}}+\,e^{i\mathbf{k}\cdot\mathbf{a_1}})$, and $C_{\mathbf{k}}=-i J_K(1+\,e^{-i\mathbf{k}\cdot\mathbf{a_3}})$.
\end{widetext}
The quadratic fermionic Hamiltonian (\ref{H_matrix})
 can be diagonalized via a standard procedure. Since the hyperhoneycomb lattice has four sites per
unit cell, the resulting band structure has four fermion
bands shown in Fig.~\ref{fig:ebandstruct}.  Note that  due to the particle-hole redundancy of the   Kitaev model in terms of Majoranas, 
the physical  excited states are described  by 
 the two positive bands only. Furthermore, the lowest positive band displays a nodal line structure \cite{OBrien2016,ybk2015prl}.

\section{Group theoretical analysis  of   lattice vibrations  on the hyperhoneycomb lattice}

\subsection{Counting Raman-Active Phonon Modes
}\label{subsec:factorGroup}

To count the Raman active phonon modes in the hyperhoneycomb lattice, we start by considering the symmetry properties and the number of atoms in the unit cell (12 Oxygens + 4 Iridium + 8 Lithium ions).  The space group for the hyperhoneycomb lattice is the nonsymmorphic $Fddd$ group no. 70. The factor group of the space group over the translation subgroup $Fddd \setminus T$ is isomorphic to the point group $D_{2h}$ \cite{Melnikov2022}. Since the Raman tensors only depend on the point group, we classify the Raman active phonon modes according to the irreps of $D_{2h}$.  The presence of inversion symmetry divides the phonons into 36 odd infrared active modes ($A_u,B_{1u},B_{2u},B_{3u}$) and 36 even Raman active modes ($A_g,B_{1g},B_{2g},B_{3g}$).

  Next we recall that $D_{2h}$ can be decomposed as a direct product of $D_2\otimes C_i$ since each of the group operations of both groups commute with each other.  For a 3D representation, the generator matrices for ${D_{2h}}={D_2\otimes C_i}$ are given by $\mathbf{x}$=diag\{1,-1,-1\}, $\mathbf{y}$=diag\{-1,1,-1\} and $\mathbf{i}$=diag\{-1,-1,-1\}. These generators can be identified with the group elements $C_{2a}$, $C_{2b} $ and inversion respectively.
 
 We choose the origin to be placed at the inversion center (origin choice 2 \cite{ITA}). 
 With respect to this origin,  we perform the group generator operations $C_{2a}$, $C_{2b} $, and inversion and store the transformation of various sites under each symmetry operation in a matrix, which we refer to as the site transformation matrix. This is a $24\times 24$ matrix (12 Oxygen + 4 Iridium + 8 Lithium ions). The direct product of the 3D generator matrices with the site transformation matrix gives the $72\times72$-dimensional matrix representations of the three generators $C_{2a}$, $C_{2b} $, and $i$. 
Using these matrices for the generators and the group multiplication table, we obtain $72\times72$ matrix representations for all 8 symmetry operations 
 of  $D_{2h}$ ($E$, $C_{2a}$, $C_{2b}$, $C_{2c}$, $i$,  $\sigma_{bc}$, $ \sigma_{ac}$, and  $\sigma_{ab}$)
in the orthorhombic coordinates $(a,b,c)$. We calculate the character $\chi$ of each group element in this naive $72\times72$ representation and get :
\[
     \begin{array}{c|cccccccc}
          & E & C_{2a} & C_{2b} & C_{2c} & i & \sigma_{bc} & \sigma_{ac} & \sigma_{ab}  \\ \hline
          \chi & 72 & -12 & 0 & -4 & 0 & 0 & 0 & 0\\
     \end{array}
\]
Then, we use the reduction formula to decompose the naive representation into the blocks of irreducible representations:
\begin{equation}\label{eqn:redfor}
    a_j=\frac{1}{h}\sum_R\,\chi^{(j)}(R)^*\,\chi(R).
\end{equation} 
Here $\chi^{(j)}(R)$ is the character of the group element $R$ in the $j^{th}$ irrep obtained from the character table and $\chi(R)$ is the character of group element $R$ obtained in the naive (reducible) representation.
The irreducible decomposition is then given by 
\begin{widetext}
    \begin{equation}\label{modect}
    \Gamma_\text{Li,Ir,O}=7A_g\oplus8B_{1g}\oplus11B_{2g}\oplus10B_{3g}\oplus 7A_u\oplus8B_{1u}\oplus11B_{2u}\oplus10B_{3u}.
\end{equation} 
\end{widetext}
The Raman active modes  of the unit cell are  then
 \begin{equation}
\Gamma^{\text R}_\text{Li,Ir,O}=7A_g\oplus8B_{1g}\oplus11B_{2g}\oplus10B_{3g}.
\end{equation} 

We note that this factor group analysis of the phonon modes is in agreement with the Raman active modes assigned by Wyckoff positions (WP). The WPs of all ions in $\beta-\text{Li}_2\text{IrO}_3$ are \cite{biffin2014unconventional}
\[
\begin{array}{c|ccccc}
          & \text{Ir} & \text{Li}_1 & \text{Li}_2 & \text{O}_1 & \text{O}_2 \\ \hline
           & 16g & 16g & 16g & 16e & 32h\\
     \end{array}
\] 
From the selection rule table \cite{bilbaoSAM}, the number of Raman active modes in each irrep channel are
\[\begin{array}{|c||c|c|c|c|}
		\hline
		\text{WP} & A_g & B_{1g} & B_{2g} & B_{3g} \\
		\hline
            \hline
		16e & 1 & 2 & 2 & 1 \\
		\hline
		16g & 1 & 1 & 2 & 2 \\
		\hline
            32h & 3 & 3 & 3 & 3 \\
            \hline
	\end{array}\]
For example, from the Wyckoff positions, we expect 2 $A_g$ modes corresponding to Li, 1 $A_g$ mode for Ir and 4 $A_g$ modes for O which adds up to the 7 $A_g$ modes. Similarly, for other irreps, we have
\begin{subequations}
        \begin{align}
        7A_g = 1 (\text{Ir}) + 2 (\text{Li}) + 1 (\text{O}_1) + 3 (\text{O}_2), \nonumber\\
        8B_{1g} = 1 (\text{Ir}) + 2 (\text{Li}) + 2 (\text{O}_1) + 3 (\text{O}_2), \nonumber\\
        11B_{2g} = 2 (\text{Ir}) + 4 (\text{Li}) + 2 (\text{O}_1) + 3 (\text{O}_2), \nonumber\\
        10B_{3g} = 2 (\text{Ir}) + 4 (\text{Li}) + 1 (\text{O}_1) + 3 (\text{O}_2). \nonumber
        \end{align}
\end{subequations}
This matches the factor group analysis in Eq.\eqref{modect}.

However, as  explained in Sec. \ref{sec:sp-phCoupling}, in our analysis we only focus on  iridium vibrations. Thus, we restrict the generators to $12\times12$-dimensional representation. Using these matrices for the generators and the group multiplication table, we obtain $12\times12$ matrix representations for all 8 symmetry operations in the orthorhombic coordinates $(a,b,c)$. We calculate the character $\chi$ of each group element in this naive $12\times12$ representation-
\[
     \begin{array}{c|cccccccc}
          & E & C_{2a} & C_{2b} & C_{2c} & i & \sigma_{bc} & \sigma_{ac} & \sigma_{ab}  \\ \hline
          \chi & 12 & 0 & 0 & -4 & 0 & 0 & 0 & 0\\
     \end{array}
\]
 This representation is reduced as
\begin{equation}
    \Gamma_\text{Ir}=A_g\oplus B_{1g}\oplus2B_{2g}\oplus2B_{3g}\oplus A_u\oplus B_{1u}\oplus2B_{2u}\oplus2B_{3u}
\end{equation}
 and the  Raman active  phonon modes are then classified as 
\begin{equation}\label{eqn:IrRaman}
    \Gamma^{\text R}_\text{Ir}=A_g\oplus B_{1g}\oplus2B_{2g}\oplus2B_{3g}.
\end{equation}

\subsection{Phonon eigenmode evaluation}\label{subsec:eigenmode}

The aim of this section  is to obtain  the $A_g$ phonon eigenmode, considering vibrations of Ir ions only.   We will do this
 by performing a canonical decomposition \cite{Serre1977} of the   representation $\Gamma$ obtained in Sec.~\ref{subsec:factorGroup}. Specifically, we calculate the coefficients  $u_{\Gamma,i}$ in the expansion $u_{\Gamma } = \sum_{i=1}^{12} u_{\Gamma , i} u_i$, where only vibration of the 4 Iridium atoms is considered. 
 This is carried out using the projection operator defined as:
\begin{equation}
    \mathcal{P}_j=\frac{n_j}{h}\sum_R \, \chi_j(R)^*\,\Gamma_R,
\end{equation}
where $\mathcal{P}_j=\mathcal{P}_{A_g}$ is the projection operator for the $A_g$ irrep, $h=8 $ is order of the $D_{2h}$ group, $n_j=1$ is the dimensionality of the $A_g$ irrep and $\chi_j(R)$ are the characters from the character table and $\Gamma_{R}$ is the reducible representation for each group operation $R$.  The rank  of $\mathcal{P}_{A_g}$ is 1 since there is  only 1 $A_g$ mode associated with the Ir atoms Eq.~\eqref{eqn:IrRaman}.

The basis vectors of the projection matrix constitute the symmetry-respecting phonon eigenvectors.   These basis vectors are obtained by factorizing the projection operator $\mathcal{P}_{A_g}$ by performing a singular value decomposition (SVD) as $\mathcal{P}_{A_g}=U \cdot \Lambda\cdot V^T$. Here $U$ and $V$ are real orthogonal matrices, while $\Lambda$ is a diagonal matrix with values 0 or 1. Thus, we obtain the basis vectors $u_{A_g}$ from the columns of the matrix $X_{A_g}:=U\cdot\Lambda$. Namely, we have 
 \begin{equation}
     u_{A_g} = \oplus_{i=1}^{12} u_{A_g , i}\, u_i,
 \end{equation}
 such that $u_{A_g}=(u^{\text{Ir}_1}_{A_g,x},u^{\text{Ir}_1}_{A_g,y},u^{\text{Ir}_1}_{A_g,z}u^{\text{Ir}_2}_{A_g,x},u^{\text{Ir}_2}_{A_g,y},u^{\text{Ir}_2}_{A_g,z},$
$u^{\text{Ir}_3}_{A_g,x},u^{\text{Ir}_3}_{A_g,y},u^{\text{Ir}_3}_{A_g,z},u^{\text{Ir}_4}_{A_g,x},u^{\text{Ir}_4}_{A_g,y},u^{\text{Ir}_4}_{A_g,z})$.
The appropriate linear combination of basis vector motivated by DFT results \cite{glamazda2016expt} in orthorhombic coordinates as visualized in Fig. \ref{fig:AgPhonon} is given as: 
\begin{equation}
u_{A_g}:=
X_{A_g}=\bigg(0,0,-\dfrac{1}{2},0,0,\dfrac{1}{2},0,0,-\dfrac{1}{2},0,0,\dfrac{1}{2}\bigg).
\end{equation}

\section{ Details of the phonon polarization bubble calculation}\label{Appendix:polbub}

In this appendix, we outline the steps  of the derivation of 
Eq.~\eqref{eqn:polbubres} for the polarization bubble,  where the total contribution is separated into those from pp- and ph-processes.
The polarization bubble  $ \Pi(\tau)$ from Eq.~(\ref{eqn:polbubexpr}) can be  written as:
\begin{widetext}
\begin{equation}\label{pipptime}
    \begin{aligned}
        \Pi^{\text{pp}}(\tau)=\frac{\lambda_{A_g}^2}{N}\sum_\mathbf{k}&\bigg[g^a(-\mathbf{k},\tau)g^a(\mathbf{k},\tau) \Tilde{\Lambda}_\mathbf{k}^{23}\Tilde{\Lambda}_\mathbf{k}^{32} + g^b(-\mathbf{k},\tau)g^b(\mathbf{k},\tau) \Tilde{\Lambda}_\mathbf{k}^{14}\Tilde{\Lambda}_\mathbf{k}^{41}\\ 
        & + g^a(-\mathbf{k},\tau)g^b(\mathbf{k},\tau) \Tilde{\Lambda}_\mathbf{k}^{13}\Tilde{\Lambda}_\mathbf{k}^{31} + g^b(-\mathbf{k},\tau)g^a(\mathbf{k},\tau) \Tilde{\Lambda}_\mathbf{k}^{24}\Tilde{\Lambda}_\mathbf{k}^{42}
        \bigg],
    \end{aligned}
\end{equation}
\begin{equation}\label{piphtime}
    \begin{aligned}
        \Pi^{\text{ph}}(\tau)=\frac{\lambda_{A_g}^2}{N}\sum_\mathbf{k}&\bigg[g^a(-\mathbf{k},\tau)\Bar{g}^a(-\mathbf{k},\tau) \Tilde{\Lambda}_\mathbf{k}^{22}\Tilde{\Lambda}_\mathbf{k}^{22} + g^b(-\mathbf{k},\tau)\Bar{g}^b(-\mathbf{k},\tau) \Tilde{\Lambda}_\mathbf{k}^{11}\Tilde{\Lambda}_\mathbf{k}^{11}\\
        & + g^a(-\mathbf{k},\tau)\Bar{g}^b(-\mathbf{k},\tau) \Tilde{\Lambda}_\mathbf{k}^{21}\Tilde{\Lambda}_\mathbf{k}^{12} + g^b(-\mathbf{k},\tau)\Bar{g}^a(-\mathbf{k},\tau) \Tilde{\Lambda}_\mathbf{k}^{12}\Tilde{\Lambda}_\mathbf{k}^{21}
        \bigg],
    \end{aligned}
\end{equation}
\end{widetext}
where the fermionic Green's functions are defined as $g^\gamma(\mathbf{k},\tau)=-\braket{T_\tau \gamma(\tau)\gamma^\dagger(0)}$ and $\Bar{g}^\gamma(\mathbf{k},\tau)=-\braket{T_\tau \gamma^\dagger(\tau)\gamma(0)}$ with $\gamma=a,\,b$ representing the Bogoliubov complex fermion flavor index.
For finite temperatures, we perform a Fourier transform to the Matsubara space:
 \begin{equation}
    \begin{aligned}
        g^\gamma(\mathbf{k},i\omega_n)&=\int_0^\beta d\tau\, e^{i\omega_n\tau}\, g^\gamma(\mathbf{k},\tau),\\
        g^\gamma(\mathbf{k},\tau)&=\frac{1}{\beta}\sum_{\omega_n} e^{-i\omega_n\tau}\, g^\gamma(\mathbf{k},i\omega_n),
    \end{aligned}
\end{equation}
where $g^\gamma(\mathbf{k},i\omega_n)= (i\omega_n-\epsilon^\gamma_\mathbf{k})^{-1}$ and $\bar{g}^\gamma(\mathbf{k},i\omega_n)= (i\omega_n+\epsilon^\gamma_\mathbf{k})^{-1}$. 

The Fourier transform to  Matsubara space for the polarization bubble is given by:
 \begin{equation}
 \Pi(i\Omega_m)=\int d\tau \, e^{i\Omega_m\tau}\,\Pi(\tau).
 \end{equation}

This Fourier transform results in a convolution of the product of Green's functions appearing in Eq.~\eqref{pipptime} and \eqref{piphtime}. We can evaluate the Matsubara sum for each pair of Green's function as:
\begin{subequations}\label{eqn:matsum}
\allowdisplaybreaks
    \begin{align}
        P_{\gamma{\gamma'}}^{gg}(i\Omega_m)&:=T\sum_{\omega_n}g^\gamma(-\mathbf{k},i\omega_n)g^{\gamma'}(\mathbf{k},i\Omega_m-i\omega_n)\nonumber\\
        &=\frac{n_F(\epsilon^\gamma_\mathbf{k})-n_F(-\epsilon^{\gamma'}_\mathbf{k})}{i\Omega_m-\epsilon^\gamma_\mathbf{k}-\epsilon^{\gamma'}_\mathbf{k}},\\
        P_{\gamma{\gamma'}}^{\Bar{g}\Bar{g}}(i\Omega_m)&:=T\sum_{\omega_n}\bar{g}^\gamma(-\mathbf{k},i\Omega_m-i\omega_n)\bar{g}^{\gamma'}(\mathbf{k},i\omega_n)\nonumber\\
        &=\frac{n_F(-\epsilon^\gamma_\mathbf{k})-n_F(\epsilon^{\gamma'}_\mathbf{k})}{i\Omega_m+\epsilon^\gamma_\mathbf{k}+\epsilon^{\gamma'}_\mathbf{k}},\\
        P_{\gamma{\gamma'}}^{g\Bar{g}}(i\Omega_m)&:=T\sum_{\omega_n}g^\gamma(-\mathbf{k},i\omega_n)\bar{g}^{\gamma'}(-\mathbf{k},i\Omega_m-i\omega_n)\nonumber\\
        &=\frac{n_F(\epsilon^\gamma_\mathbf{k})-n_F(\epsilon^{\gamma'}_\mathbf{k})}{i\Omega_m-\epsilon^\gamma_\mathbf{k}+\epsilon^{\gamma'}_\mathbf{k}},
        \\
        P_{\gamma{\gamma'}}^{\Bar{g}g}(i\Omega_m)&:=T\sum_{\omega_n}\bar{g}^\gamma(\mathbf{k},i\omega_n)g^{\gamma'}(\mathbf{k},i\Omega_m-i\omega_n)\nonumber\\
        &=\frac{n_F(-\epsilon^\gamma_\mathbf{k})-n_F(-\epsilon^{\gamma'}_\mathbf{k})}{i\Omega_m+\epsilon^\gamma_\mathbf{k}-\epsilon^{\gamma'}_\mathbf{k}}.
    \end{align}
\end{subequations}

To obtain real frequencies, we perform an analytic continuation by substituting $i\Omega_m\rightarrow\Omega+i\delta$.

\begin{figure}[t]
    \centering
    \includegraphics[width=0.85\linewidth]{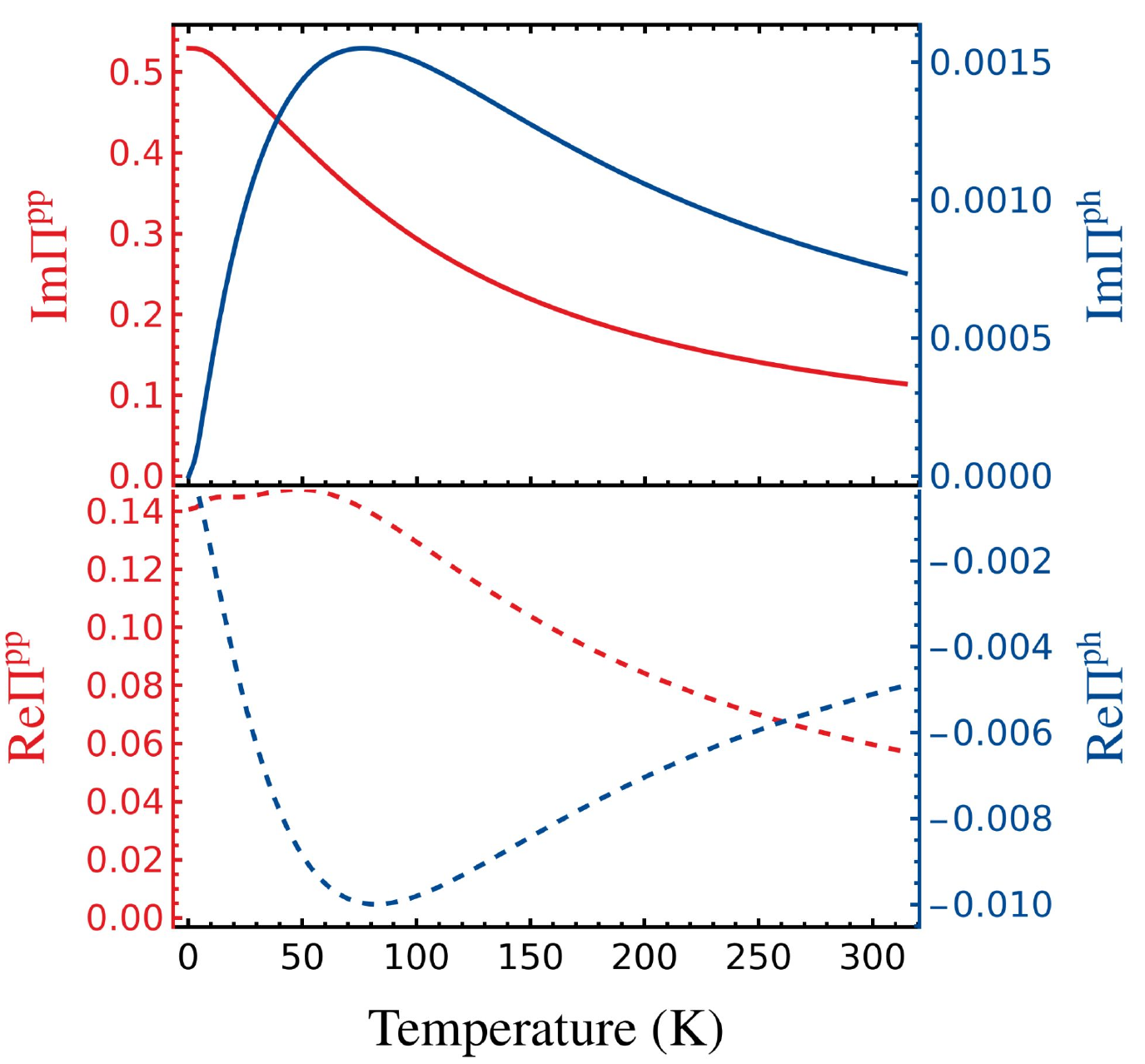}
    \caption{Temperature evolution of the real and imaginary parts of the polarization bubble, Re$\Pi(\Omega, T)$ and Im$\Pi(\Omega, T)$, at a fixed frequency of $\Omega_{A_g}=24$ meV. The pp- and ph-channels are shown separately.}
    \label{fig:pitmpplt}
\end{figure}

Finally, we comment on the temperature evolution of pp- and ph-channels of the polarization bubble as shown in Fig. \ref{fig:pitmpplt}. We observe that the pp-channel is dominant compared to ph-channel. This is because an incoming phonon with sufficient energy can decay into a pair of fermions. The kinematic constraints for this process can be satisfied in more ways, especially due to the presence of two physical bands leading to a dominant pp-channel. On the other hand, in a ph-process, a phonon scatters off a fermion and excites it to a higher energy state by transferring the energy of the optical phonon. This process requires a finite population of the fermionic states which only occurs at finite temperatures. Thus, both Re$\Pi$ and Im$\Pi$ vanish at zero temperature in the ph-channel.

Furthermore, the renormalized phonon peak  \eqref{eq:broadened_phonon} parameters such as peak position and peak width are controlled by Re$\Pi$ and Im$\Pi$ respectively. As noted above, the total contribution obtained by summing over pp- and ph-channels is expected to be pp-dominant. We find that the renormalization to the peak position decreases with temperature as seen from a decrease in Re$\Pi$ shown by red dashed curve in Fig. \ref{fig:pitmpplt}. This is expected because the spins are not fractionalized at higher temperatures and there are no Majorana fermions for the phonons to couple. The phonon peak width renormalization exhibits a similar decreasing behavior as seen by the decay of Im$\Pi$ with temperature denoted by red curve in Fig. \ref{fig:pitmpplt}. Thus, with increasing temperature, the effects of the spin-phonon coupling die out.

\section{Magnetic Raman response}\label{app:mag-raman}

 In momentum space within the Bogoliubov basis, the magnetic Raman operator (\ref{RamanLF}) can be expressed as follows:
\begin{equation}
 \mathcal{R}_{\text{em-s}}=\nu\sum_\mathbf{k}A_\mathbf{k}^\dagger\Tilde{R}_\mathbf{k}A_\mathbf{k},
 \end{equation}
where $\nu$ is the photon-spin coupling constant,
$\Tilde{R}_\mathbf{k}=U_\mathbf{k}^\dagger R_{\mathbf{k}} U_\mathbf{k}$, and  $U_\mathbf{k}$ 
 is the unitary matrix that diagonalizes the spin Hamiltonian  \eqref{eqn:c2a}. 
 The matrix $R_{\mathbf{k}}$ is given by
\begin{equation}
    R_\mathbf{k}=\begin{pmatrix}
    0 & 0 & \mathcal{B}_{\mathbf{k}} & \mathcal{A}_{\mathbf{k}} \\
    0 & 0 & \mathcal{A}_{\mathbf{k}} & \mathcal{C}_{\mathbf{k}} \\
    \mathcal{B}^*_{\mathbf{k}} & \mathcal{A}^*_{\mathbf{k}} & 0 & 0 \\
    \mathcal{A}^*_{\mathbf{k}} & \mathcal{C}^*_{\mathbf{k}} & 0 & 0 
\end{pmatrix},
\end{equation}
with $\mathcal{A}_{\mathbf{k}}=-iJ_KP_z$, $\mathcal{B}_{\mathbf{k}}=-i J_K(P_{x'}\,e^{i\mathbf{k}\cdot\mathbf{a_2}}+P_y\,e^{i\mathbf{k}\cdot\mathbf{a_1}})$, $\mathcal{C}_{\mathbf{k}}=-i J_K( P_x+ P_{y'}\,e^{-i\,\mathbf{k}\cdot\mathbf{a_3}})$,
where we define the polarization factors $P_t=(\mathbf{M}_t\cdot\boldsymbol{\varepsilon}_{\text{in}})(\mathbf{M}_t\cdot\boldsymbol{\varepsilon}_{\text{out}})$ for the nearest neighbor bonds indexed by $t$.
The Raman intensity is then computed as 
\begin{equation}
I_{\text{em-s}}(i\Omega_m)=  -\int_0^\beta  d\tau e^{i\Omega_m \tau}\braket{T_{\tau}\mathcal{R}_{\text{em-s}}(\tau)\mathcal{R}_{\text{em-s}}(0)}.
\end{equation}

We use the same method for calculating this correlator as we did for the polarization bubble in Appendix
\ref{Appendix:polbub}. By expressing the Raman operator in  the Bogoliubov basis, we combine it with the Green's functions describing Majorana fermion's propagators. After performing the Matsubara summation, the result can be written as:
 
    \begin{align}  \label{eqn:Iems-pp}
        &I_{\text{em-s}}^{\text{pp}}(i\Omega_m)= \frac{\nu^2}{N}\sum_\mathbf{k} \bigg[P^{gg}_{aa}(i\Omega_m)\Tilde{R}_\mathbf{k}^{23}\Tilde{R}_\mathbf{k}^{32}   +  \\ \nonumber
  &P^{gg}_{bb}(i\Omega_m)\Tilde{R}_\mathbf{k}^{14}\Tilde{R}_\mathbf{k}^{41}         +P^{gg}_{ba}(i\Omega_m)\Tilde{R}_\mathbf{k}^{13}\Tilde{R}_\mathbf{k}^{31}  +P^{gg}_{ab}(i\Omega_m)\Tilde{R}_\mathbf{k}^{42}\Tilde{R}_\mathbf{k}^{24}
        \bigg],
\end{align}
        
 \begin{align} 
    \label{eqn:Iems-ph}
        &I_{\text{em-s}}^{\text{ph}}(i\Omega_m)= \frac{\nu^2}{N}\sum_\mathbf{k} \Tilde{R}_\mathbf{k}^{43}\Tilde{R}_\mathbf{k}^{34}\bigg[P^{g\bar{g}}_{ab}(i\Omega_m)
        + P^{g\bar{g}}_{ba}(i\Omega_m)
        \bigg].
        \end{align}
As noted earlier, the contribution of the ph-processes
to the Raman intensity arises only due to the presence of two distinct physical bands, $a$ and $b$ in the Majorana  fermion spectrum  in Kitaev hyperhoneycomb model. This is in stark contrast to the 2D case, where the ph-processes  contribution to the Raman response is not allowed.

 Finally, the Raman susceptibility 
 that characterizes the response of a material to incident light in a Raman scattering experiment at a given frequency
 is evaluated as
 $\chi(\Omega)=I_{\text{em-s}}(i\Omega_m)|_{i\Omega_m\rightarrow \Omega +i\delta}$. The Raman intensity is related to the imaginary part of the Raman susceptibility through the fluctuation-dissipation theorem. The relationship is given by:
 \cite{Coleman_2015}:
\begin{equation}
    I(\Omega)=-\frac{2}{1-e^{-\beta\Omega}}\;\mathrm{ Im}\chi(\Omega),
\end{equation}
where $\beta=1/T$
(with $k_B=1$ being the Boltzmann constant). This formula ensures that the intensity is correctly scaled  with temperature according to the Bose-Einstein distribution.\\

\begin{figure*}
    \centering
    \includegraphics[width=\linewidth]{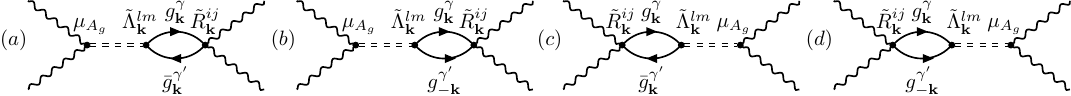}
    \caption{Feynman diagrams contributing to the Fano lineshape: (a) and (c) show the contributions from the  ph-channel and (b) and (d) show the contributions from the  pp-channel.  The indices $i,j,l,m \in (1,4)$. The indices $\gamma,\gamma'$ can take values $a$ or $b$, representing the flavor of the complex fermions. The specific values of $\gamma$ and $\gamma'$ are determined by the choice of $i,j,l,m$.}
    \label{fig:fanoppph}
\end{figure*}

\section{ Evaluation of the Fano Intensity}\label{App:Fano}
The lowest (first) order diagrams that contribute to the Fano lineshape are given by the following correlation functions:
\begin{subequations}
    \begin{align}
        I_L^{(1)}(\tau)=\int_0^\beta\, d\tau_1 \,\braket{T_\tau\mathcal{R}^{A_g}_{\text{em-ph}}(\tau)\mathcal{R}^{A_g}_{\text{em-s}}(0)\mathcal{H}^{A_g}_{\text{s-ph}}(\tau_1)}, \\
        I_R^{(1)}(\tau)=\int_0^\beta\, d\tau_1 \,\braket{T_\tau\mathcal{R}^{A_g}_{\text{em-s}}(\tau)\mathcal{R}^{A_g}_{\text{em-ph}}(0)\mathcal{H}^{A_g}_{\text{s-ph}}(\tau_1)}. 
    \end{align}
\end{subequations}
Plugging in the Raman vertices, coupling vertex, and performing Wick's contraction to form propagators from the product of operators yields Eq. \eqref{eqn:IlIr}, 
\begin{widetext}
\begin{subequations}\label{eqn:IlIr}
    \begin{align}
        I_L^{(1)}(\tau)&=\int_0^\beta\, d\tau_1 \,\braket{T_\tau\bigg(\mu_{A_g}\mathcal{R}_{A_g}u_{A_g} \bigg)(\tau)\bigg(\nu\sum_\mathbf{k}A_\mathbf{k}^\dagger\Tilde{R}_\mathbf{k}A_\mathbf{k}\bigg)(0)\bigg(\lambda_{A_g}u_{A_g} \sum_\mathbf{k'}A_\mathbf{k'}^\dagger\Tilde{\Lambda}_\mathbf{k'}A_\mathbf{k'}\bigg)(\tau_1)}\\
&=\nu\lambda_{A_g}\mu_{A_g} R^{cc}_{A_g}\int_0^\beta\, d\tau_1 \sum_\mathbf{k}
        \braket{T_\tau u_{A_g}(\tau)u_{A_g}(\tau_1)}\braket{T_\tau (A_\mathbf{k}^\dagger)^i(0)A^m_\mathbf{k}(\tau_1)}
    \braket{T_\tau A^j_\mathbf{k}(0)(A_\mathbf{k}^\dagger)^l (\tau_1)} \Tilde{R}^{ij}_\mathbf{k}\Tilde{\Lambda}^{lm}_\mathbf{k},\nonumber\\
I_R^{(1)}(\tau)&=\int_0^\beta\, d\tau_1 \,\braket{T_\tau\bigg(\nu\sum_\mathbf{k}A_\mathbf{k}^\dagger\Tilde{R}_\mathbf{k}A_\mathbf{k}\bigg)(\tau)\bigg(\mu_{A_g}\mathcal{R}_{A_g}u_{A_g} \bigg)(0)\bigg(\lambda_{A_g}u_{A_g} \sum_\mathbf{k'}A_\mathbf{k'}^\dagger\Tilde{\Lambda}_\mathbf{k'}A_\mathbf{k'}\bigg)(\tau_1)}\\
&=\nu\lambda_{A_g}\mu_{A_g} R^{cc}_{A_g}\int_0^\beta\, d\tau_1 \sum_\mathbf{k}
        \braket{T_\tau (A_\mathbf{k}^\dagger)^i(\tau)A^m_\mathbf{k}(\tau_1)}
    \braket{T_\tau A^j_\mathbf{k}(\tau)(A_\mathbf{k}^\dagger)^l (\tau_1)} \braket{T_\tau u_{A_g}(0)u_{A_g}(\tau_1)}\Tilde{R}^{ij}_\mathbf{k}\Tilde{\Lambda}^{lm}_\mathbf{k}.\nonumber
    \end{align}
\end{subequations}
\end{widetext}

By performing a Fourier transform to Matsubara frequency space and subsequently carrying out the Matsubara sums, we derive the final expressions for the Fano intensity, as given in Eq.~\eqref{eqn:fano}. 
The left and right Fano diagrams, with the ph- and pp-channels explicitly shown, are presented in Fig.~\ref{fig:fanoppph}. Their explicit expressions are 
\begin{widetext}
\begin{subequations}\label{eqn:fano}
\begin{equation}\label{eqn:ilpp}
    \begin{aligned}
        I_L^{\text{pp}}(i\Omega_m)=-\frac{\lambda_{A_g}\nu\mu_{A_g}}{N}\sum_\mathbf{k} \mathcal{D}(i\Omega_m)\bigg[P_{aa}^{gg}(i\Omega_m)\Tilde{\Lambda}_\mathbf{k}^{23}\Tilde{R}_\mathbf{k}^{32} + P_{bb}^{gg}(i\Omega_m)\Tilde{\Lambda}_\mathbf{k}^{14}\Tilde{R}_\mathbf{k}^{41}
         + P_{ba}^{gg}(i\Omega_m) \Tilde{\Lambda}_\mathbf{k}^{13}\Tilde{R}_\mathbf{k}^{31} + P_{ab}^{gg}(i\Omega_m)\Tilde{\Lambda}_\mathbf{k}^{24}\Tilde{R}_\mathbf{k}^{42}
        \bigg],
    \end{aligned}
\end{equation}
    \begin{equation}\label{eqn:irpp}
    \begin{aligned}
        I_R^{\text{pp}}(i\Omega_m)=-\frac{\lambda_{A_g}\nu\mu_{A_g}}{N}\sum_\mathbf{k} \mathcal{D}(i\Omega_m)\bigg[P_{aa}^{gg}(i\Omega_m)\Tilde{R}_\mathbf{k}^{23}\Tilde{\Lambda}_\mathbf{k}^{32} + P_{bb}^{gg}(i\Omega_m)\Tilde{R}_\mathbf{k}^{14}\Tilde{\Lambda}_\mathbf{k}^{41}
        + P_{ba}^{gg}(i\Omega_m) \Tilde{R}_\mathbf{k}^{13}\Tilde{\Lambda}_\mathbf{k}^{31} + P_{ab}^{gg}(i\Omega_m)\Tilde{R}_\mathbf{k}^{24}\Tilde{\Lambda}_\mathbf{k}^{42}
        \bigg],
    \end{aligned}
\end{equation}
\begin{equation}\label{eqn:ilph}
        \begin{aligned}
            I_L^{\text{ph}}(i\Omega_m)=-\frac{\lambda_{A_g}\nu\mu_{A_g}}{N}\sum_\mathbf{k} \mathcal{D}(i\Omega_m)\bigg[P_{ab}^{\Bar{g}g}(i\Omega_m)\Tilde{\Lambda}_\mathbf{k}^{34}\Tilde{R}_\mathbf{k}^{43} + P_{ba}^{g\bar{g}}(i\Omega_m)\Tilde{R}_\mathbf{k}^{12}\Tilde{\Lambda}_\mathbf{k}^{21}
        \bigg],
        \end{aligned}
\end{equation}
\begin{equation}\label{eqn:irph}
        I_R^{\text{ph}}(i\Omega_m)=-\frac{\lambda_{A_g}\nu\mu_{A_g}}{N}\sum_\mathbf{k} \mathcal{D}(i\Omega_m) P_{ba}^{g\Bar{g}}(i\Omega_m) \bigg[\Tilde{R}_\mathbf{k}^{12}\Tilde{\Lambda}_\mathbf{k}^{21} + \Tilde{R}_\mathbf{k}^{34}\Tilde{\Lambda}_\mathbf{k}^{43}
        \bigg],
\end{equation}
\end{subequations}
\end{widetext}
where $\mathcal{D}(i\Omega_m)$  is the phonon propagator  renormalized via the  usual Dyson equation.

\bibliography{refs.bib}
\bibliographystyle{apsrev4-1}
\end{document}